\documentclass[9pt,twocolumn,twoside]{osajnl}

\journal{josaa} 

\usepackage{amsmath}
\usepackage{amssymb}
\usepackage{bm}
\usepackage{float}
\usepackage{graphicx}	
\usepackage{listings}

\newcommand{\eref}[1]{Eq.(\ref{#1})}
\newcommand{\fref}[1]{Fig.\ref{#1}}
\newcommand{\tref}[1]{Table \ref{#1}}

\newcommand{\average}[1]{\left \langle {#1} \right \rangle}
\newcommand{\T}{\mathsf{T}} 
 
\newcommand{\alphavec}{{\boldsymbol{\alpha}}} 
 
\newcommand{\hlone}[1]{#1}
\newcommand{\hltwo}[1]{#1}

\setboolean{shortarticle}{false} 

\title{Fast Iterative Tomographic Wave-front Estimation with Recursive Toeplitz Reconstructor Structure for Large Scale Systems}

\author[1,2]{Yoshito H. Ono}
\author[2]{Carlos Correia}
\author[3]{Rodolphe Conan}
\author[4]{Leonardo Blanco}
\author[2]{Benoit Neichel}
\author[2,4]{Thierry Fusco}

\affil[1]{Subaru Telescope, National Astronomical Observatory of Japan, 650 North A'ohoku Place, Hilo, HI 96720, U.S.A.}
\affil[2]{Aix Marseille Univ, CNRS, LAM, Laboratoire d'Astrophysique de Marseille, Marseille, France}
\affil[3]{GMTO Corporation, 465 N. Halstead Street, Suite 250. Pasadena, CA 91107}
\affil[4]{ONERA, the French Aerospace Laboratory, F-92322 Chatillon, France}

\affil[*]{Corresponding author: ono@naoj.org}

\ociscodes{(010.1080) Active or adaptive optics; (100.3190) Inverse problems}

\begin{abstract}
Tomographic wave-front reconstruction is the main computational bottleneck to realize real-time correction for turbulence-induced wave-front aberrations in future laser-assisted tomographic adaptive-optics (AO) systems for ground-based Giant Segmented Mirror Telescopes (GSMT), because of its unprecedented number of degrees of freedom, $N$, \hlone{i.e. the number of measurements from wave-front sensors (WFS)}. In this paper, we provide an efficient implementation of the minimum-mean-square error (MMSE) tomographic wave-front reconstruction mainly useful for some classes of AO systems not requiring a multi-conjugation, such as laser-tomographic AO (LTAO), multi-objcet AO (MOAO) and ground-layer AO (GLAO) systems, \hltwo{but also applicable to multi-conjugate AO (MCAO) systems}. This work expands that by R. Conan [ProcSPIE, 9148, 91480R (2014)] to the multi-wave-front, tomographic case using natural and laser guide stars. The new implementation exploits the Toeplitz structure of covariance matrices used in a MMSE reconstructor, which leads to an overall $O(N\log N)$ real-time complexity compared to $O(N^2)$ of the original implementation using straight vector-matrix multiplication. We show that the Toeplitz-based algorithm leads to 60\,nm rms wave-front error improvement for the European Extremely Large Telescope Laser-Tomography AO system over a well-known sparse-based tomographic reconstruction, \hltwo{but the number of iterations required for suitable performance is still beyond what a real-time system can accommodate to keep up with the time-varying turbulence}.

\end{abstract}

\setboolean{displaycopyright}{true}

\begin{document}

\maketitle

\section{Introduction}
\label{sec:introduction}
The tomographic wave-front reconstruction (WFR) in adaptive-optics (AO) systems using multiple guide-stars (GS) and wave-front sensors (WFS) represents a computational challenge for real-time atmospheric turbulence correction at a few hundreds to thousands Hertz frame-rates. This is especially so for AO systems in future giant segmented mirror telescopes (GSMT) with primary diameters in the 20\,m--40\,m range, because of its unprecedented number of degrees of freedom, $N$, \hlone{i.e. the number of measurements from WFSs}. A flurry of methods has been developed, providing reduced complexity algorithms that could accelerate simulation of large systems and later be mapped onto real-time computers \cite{poyneer05,bond17,rosensteiner11,ramlau14,ellerbroek09,thiebaut10}. 


The tomographic WFR problem can be divided into two steps: i) the estimation of the wave-front in the pupil plane along GS directions and ii) the three-dimensional estimation in the turbulence volume from its projections. Efficient methods to solve for the first step are for instance the Fourier-domain reconstructor \cite{poyneer05,bond17} and more recently \cite{rosensteiner11} that promise to decrease the computational complexity from its original $O(N^2)$ to respectively $O(N\,log(N))$ and $O(N)$. Work reviewed in Ramlau et al \cite{ramlau14} falls under this category. 

A somewhat different path has been followed by others in that an explicit minimum-mean-square error (MMSE) residual cost-functional is solved for leading to formulations that are amenable to sparse representations under some reasonable approximations (and therefore to efficient computational
reconstruction methods). The most representative examples are reviewed in Ellerbroek et al \cite{ellerbroek09}, with some more additions along the same lines from \cite{thiebaut10} using a fractal approximation of the regularizing stratified phase covariance term. 

More recently, for some classes of AO systems not requiring multi-conjugation, it has been noted that the MMSE reconstruction can be further simplified if we skip the explicit estimation of the 3D wave-front profiles to estimate instead the pupil-plane wave-front in the directions of interest only \cite{Vidal-10, Martin-12, correia14, Conan-14} (hereinafter, referred to as \textit{spatio-angular} WFR), which is suitable for multi-object AO (MOAO), laser-tomography AO (LTAO) and optionally ground-layer AO (GLAO) systems. The spatio-angular WFR doesn't require any approximation and thus can provide more accurate estimation than the sparse reconstructor. However, for large-scale systems, this explicit formulation requires instantiating huge covariance matrices with the overall complexity remaining O($N^2$) which is still a cause of computational bottleneck. 

An efficient implementation of the \textit{spatio-angular} WFR is proposed by  R. Conan \cite{Conan-14} for a natural GS (NGS) based classical single-conjugate AO system (SCAO). This implementation reduces the computational complexity to $O(N\,log(N))$ from its original $O(N^2)$ by exploiting the Toeplitz structure of the covariance matrices. In this paper we generalize this method to the tomographic system using NGSs and laser GSs (LGS). Especially for tomographic systems with multiple LGSs, we develop a way to deal with the spatial sampling change at high altitudes due to the cone effect of LGSs and \hltwo{the removal of low-order modes not measurable by LGSs}, whilst keeping the Toeplitz structure of the covariance matrices. \hltwo{In passing, although this is not the main motivation of this paper, in doing so we can show that MCAO systems are also covered by the implementation presented here}.




This paper is organized as follows. In \S \ref{sec:tomoWRF} we review the reconstruction formulations and the fast implementations they are amenable to when considering large systems for GSMT. We give a thorough account of the development of a spatio-angular WFR algorithm exploiting the Toeplitz block structure of the reconstructor matrices. In \S \ref{sec:perfComparison} we compare performance on 8\,m class telescopes by analytic and Monte-Carlo, physical-optics simulations and extend to the European Extremely Large Telescope in \S \ref{sec:eltHarmoni}. In \S \ref{sec:realtime}, the real time readiness of the Toeplitz algorithm using parallel computing with graphical processing unit (GPU) is discussed. Final discussion and remarks are laid out in \S \ref{sec:conclusion}.

\section{Tomographic Wave-front Reconstruction}
\label{sec:tomoWRF}
\begin{figure}
    \centering
    \includegraphics[bb=0 0 486 405, width=0.4\textwidth]{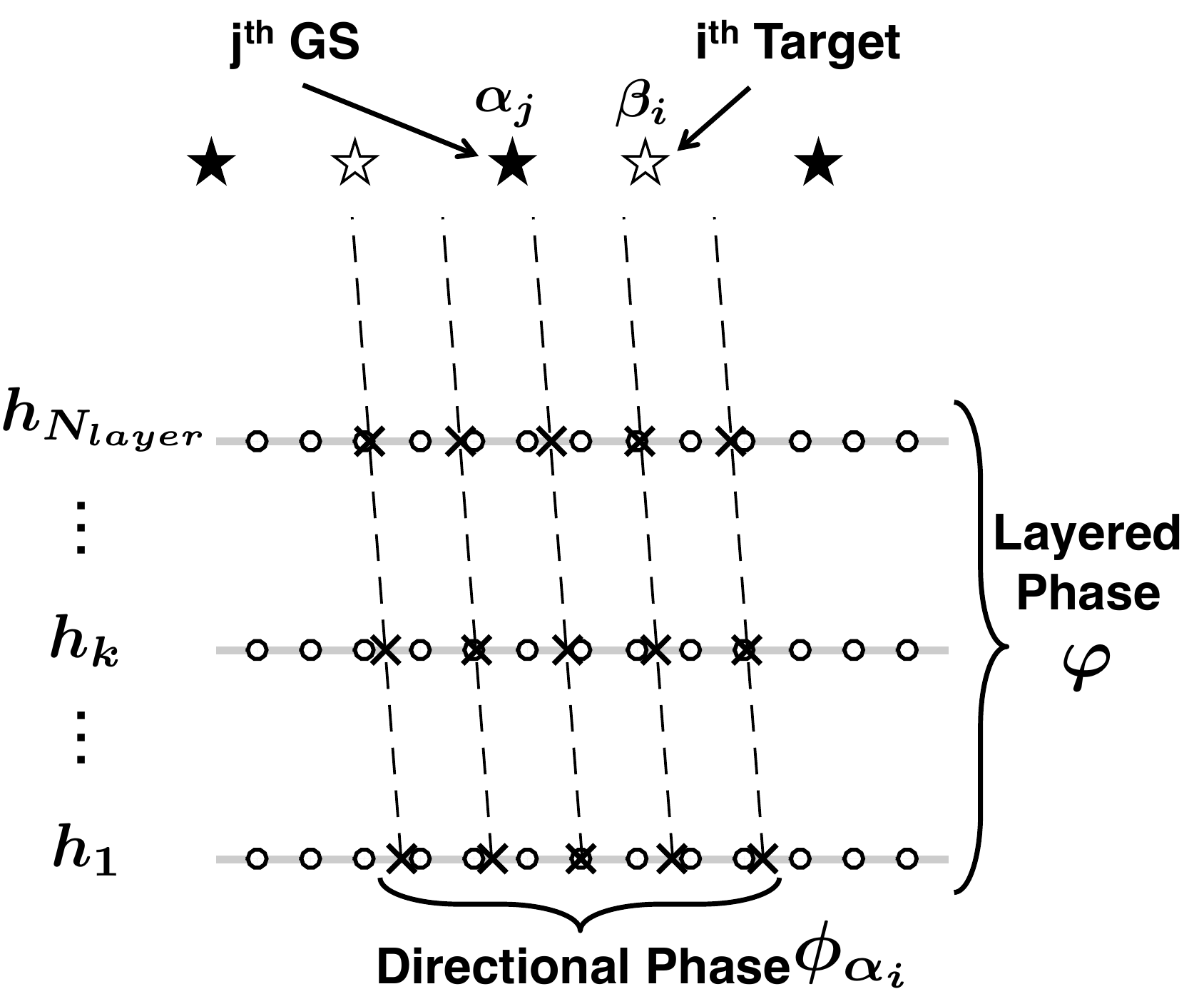}
    \caption{Schematic figure of the assumed situation with $N_{layer}$ turbulence layers, $N_{gs}$ GSs and $N_{target}$ targets. The layered phase grids of $N_{layer}$ turbulence layers are represented by the circles. The cross symbols show the directional phase grid in a GS direction $\bm{\alpha_j}$. The target direction is denoted by $\bm{\beta_i}$.}
    \label{fig:intro}
\end{figure}

We assume a situation shown in \fref{fig:intro}, where the atmospheric turbulence is expressed as $N_{layer}$ thin turbulent layers at multiple altitudes $h_k\ (k=1,\cdots,N_{layer})$, and the phase aberration on the atmospheric layer grids (circle symbols in \fref{fig:intro}) is noted as $\bm{\varphi}$. This term is referred to as the \textit{layered phase}. The tomographic computation is made based on $N_{gs}$ GSs in directions of $\bm{\alpha_j}=(\alpha_j^x,\alpha_j^y)\ (j=1,\cdots,N_{gs})$ using measurements from $N_{gs}$ WFSs with DM corrections applied over \hlone{$N_{target}$ target directions $\bm{\beta_i}\ (i=1,\cdots,N_{target})$}. The \textit{directional phase} $\bm{\phi_{\alpha_j}}$ is integrated on the pupil-plane by ray-tracing the layered contributions $\bm{\varphi}$ and interpolating at the intercepts. In other words, the phase values of the propagating wave-fronts (cross symbols in \fref{fig:intro}) are interpolated from $\bm{\varphi}$ using a \hlone{bi-linear interpolation matrix $\bm{P_{\alpha_j,k}}$} at $h_k$ \cite{gilles08a}, and the final wave-front phase at pupil-plane is given by an integration over altitudes, i.e.
\begin{equation}
    \bm{\phi_{\alpha_j}}=\bm{P_{\alpha_j}}\bm{\varphi}=\sum_{k=1}^{N_{layer}}\bm{P_{\alpha_j,k}}\bm{\varphi},
\end{equation}

WFS measurements are modeled by a linear WFS operator $\bm{G}$ and a noise vector $\bm{\eta}$ as
\begin{equation}\label{eq:linearModel}
    \bm{s_{\alpha_i,\eta}} = \bm{G P_{\alpha_i}}\bm{\varphi} + \bm{\eta},
\end{equation}
where $\bm{s_{\alpha_i}}=\bm{G P_{\alpha_i}}\bm{\varphi}$ is denoted as the noiseless measurement component. The concatenation of measurements from all $N_{gs}$ WFSs is denoted by $\bm{s_{\alpha,\eta}}$. 

In the remainder of the paper, variables with a hat symbol represent an estimated quantity, $\bm{\Sigma_{xy}}$ represents auto- and cross-covariance matrices of the vectors indicated in subscript, i.e. $\bm{\Sigma_{xy}}=\average{\bm{x}\bm{y}}$ where \hlone{$\average{\cdot}$ stands for ensemble average over time}.

\subsection{Minimum Mean Square Error reconstruction}
We will address MMSE reconstructors which minimize the aperture-plane residual wave-front error in a single direction which is given by \hlone{the Euclidean norm $L_2$} over the telescope pupil $\Omega$ of the difference between the input and the correction phases
\begin{equation}
\label{eq:criteria}
    \sigma^2_{\bm{\beta_i}} = \left|\left|
    \bm{\phi_{\bm{\beta_i}}}-\bm{\widehat{\phi}_{\beta_i}}
    \right|\right|^2_{L_2(\Omega)}.
\end{equation}
where the estimated directional phase $ \bm{\widehat{\phi}_{\beta_i}} = \bm{R_{\beta_i}} s_\alphavec$ with the reconstructor matrix $\bm{R_{\beta_i}}$ minimizing
\begin{equation}
\label{eq:sigma2}
	\bm{R_{\beta_i}}= \arg \min_{\bm{R_{\beta_i}}} \average{\sigma^2_{\bm{\beta_i}}}.
\end{equation}
for the direction of optimization $\bm{\beta_i}$ by solving for  $\partial \sigma^2/\partial\bm{R_{\beta_i}} = \bm{0}$. 
According to the Marechal's approximation for the Strehl ratio i.e.  $\text{SR}\geq(1-\sigma^2/2)^2$, minimizing \eref{eq:criteria} is equal to maximize image quality in terms of the Strehl ratio.


The MMSE solution was developed in \cite{fusco01, ellerbroek02} and others remaining general and applicable to multi-conjugate AO (MCAO) systems, which requires an explicit estimation of the layered phase $\bm{\varphi}$ for the multiple DM conjugation:
\begin{align}
    \label{eq:covBasedReconstructors1}
    \bm{R} & = \bm{\Sigma_{\varphi s_{\alpha}}}\left(\bm{\Sigma_{s_{\alpha}s_{\alpha}}}+
    \bm{\Sigma_{\eta\eta}}\right)^{-1}\\
    \label{eq:covBasedReconstructors2}
    & = \bm{\Sigma_{\varphi \varphi}}\bm{P_\alpha}^\T\bm{G}^\T\left(\bm{GP_\alpha}\bm{\Sigma_{\varphi \varphi}}\bm{P_\alpha}^\T\bm{G}^\T+
    \bm{\Sigma_{\eta\eta}}\right)^{-1}\\
    \label{eq:covBasedReconstructors3}
    & = (\bm{P_\alpha}^T\bm{G}^T\bm{\Sigma_{\eta\eta}}^{-1}\bm{G}\bm{P_\alpha}
    +\bm{\Sigma_{\varphi\varphi}}^{-1})^{-1}\bm{P_\alpha}^T\bm{G}^T\bm{\Sigma_{\eta\eta}}^{-1}.
\end{align}
 The reconstructor for the directional phase in \eref{eq:sigma2} is given by $\bm{R_{\beta_i}}=\bm{P_{\beta_i}}\bm{R}$. Although the WFR is followed by the fitting process to determine commands sent to DM(s) \cite{gilles08a}, in this paper we focus only on the WFR process.

\hltwo{For LGSs-based systems the removal of low-order modes (tip/tilt/focus) needs be taken into account in the reconstructor because these modes are not measurable by LGSs. We will address this point later in \S \ref{sec:tomoWRF}.}

\subsection{Sparse reconstructor formulations}
It has early been recognized that the formulation of \eref{eq:covBasedReconstructors3} is amenable to a sparse representation in an attempt to reduce the computational burden with iterative implementations \cite{ellerbroek02} since the number of operations in a matrix-vector multiplication (MVM) with a sparse matrix is proportional to the number of non-zero elements in the matrix. The directional phase is estimated iteratively using the formulation of \eref{eq:covBasedReconstructors3} divided into the following three steps:
\begin{equation}
\label{eq:sparse1}
    \bm{\zeta}=\bm{P_\alpha}^T\bm{G}^T\bm{\Sigma_{\eta\eta}}^{-1}\bm{s_{\alpha_i,\eta}}
\end{equation}
and a layered phase estimation $\bm{\widehat{\varphi}}$ with an iterative method
\begin{equation}
\label{eq:sparse2}
    (\bm{P_\alpha}^T\bm{G}^T\bm{\Sigma_{\eta\eta}}^{-1}\bm{G}\bm{P_\alpha}
    +\bm{\Sigma_{\varphi\varphi}}^{-1})\bm{\widehat{\varphi}}=\bm{\zeta}
\end{equation}
thus avoiding inverting explicitly $(\bm{P_\alpha}^T\bm{G}^T\bm{\Sigma_{\eta\eta}}^{-1}\bm{G}\bm{P_\alpha}
    +\bm{\Sigma_{\varphi\varphi}}^{-1})$ which would cause both off-line issues to compute and store the matrix and on-line increased burden since it is a full matrix. Finally, the phase is ray-traced to the aperture
$\bm{\widehat{\phi}_{\beta_i}}=\bm{P_{\beta_i}}\bm{\widehat{\varphi}}$, using a low complexity step, since only 4 elements per phase sample per layer have non-zero values \cite{gilles08a}. 

Assuming the use of Shack-Hartmann WFS (SH-WFS) with \hlone{a subaperture size of $d$}, measurements are the spatial derivatives of the wave-front averaged on each subaperture. The linear SH-WFS operator $\bm{G}$ admits a discrete approximation from uniform $3\times3$ stencils~\cite{gilles13} given by
\begin{align}
    \text{stencil}(\bm{G})_x&=\frac{1}{2d}
    \begin{bmatrix}
    -1/4 & 0 & 1/4\\
    -1/2 & 0 & 1/2\\
    -1/4 & 0 & 1/4
    \end{bmatrix}\\
    \text{stencil}(\bm{G})_y&=\text{stencil}(\bm{G})_x^T
\end{align}
and therefore has only 6 non-zero elements per subaperture.  The noise covariance matrix is generally considered as a diagonal matrix by assuming a zero-mean additive Gaussian noise. In the LGSs case, although $x$- and $y$-measurements are correlated due to the spot elongation, 
only 3 central diagonals have non-zero values \cite{robert10}. The inverse covariance matrix of the layered phase $\bm{\Sigma_{\varphi\varphi}}^{-1}$ is a dense matrix but it can be approximated as a sparse matrix with a discrete Laplacian operator $\bm{L}$ as $\bm{\Sigma_{\varphi\varphi}}^{-1}\approx\bm{L}^T\bm{L}$~\cite{ellerbroek02}.

\subsection{Spatio-angular tomographic formulation}
We now enter the core matter of this paper. We start from the formulation in \eref{eq:covBasedReconstructors1} to estimate the directional phase. The directional phase $\bm{\phi_{\beta_i}}$ is directly connected with the measurements $\bm{s_{\alpha,\eta}}$ through the covariance matrix $\bm{\Sigma_{\phi_{\beta_i} s_{\alpha}}}=\bm{P_{\beta_i}}\bm{\Sigma_{\varphi s_{\alpha}}}$, and the explicit estimation of the layered phase is circumvented, \hltwo{thus allowing us to make the size of the reconstruction matrix compact}.

The required covariance matrices are derived theoretically along with the measurement model. The slope-slope and phase-slope covariance matrices are computed through numerical integration~\cite{Butterley-06} or the fast Fourier transform (FFT) ~\cite{Vidal-10}. Approximated measurement models are also proposed for the fast computation of the covariance matrices~\cite{Martin-12}. In this paper, we investigate three measurement models to derive the theoretical covariance matrices: the accurate FFT model, Hudgin-like model and Fried model. The details of the models are summarized in Appendix. \hltwo{The Hudgin-like model and Fried model define a measurement with two and four discrete phase points, respectively. Although models defining a measurement with more than 4 points would give better approximations, these models result in more computations and loose their advantage for a fast computation. Therefore, we only focus on the Hudgin-like model and Fried model as fast approximated gradient models in this paper.}

The iterative implementation of \eref{eq:covBasedReconstructors1} is given by the following two steps:
\begin{equation}
\label{eq:covIterative1}
    \left(\bm{\Sigma_{s_{\alpha}s_{\alpha}}}+\bm{\Sigma_{\eta\eta}}\right)\bm{\zeta}=\bm{s_{\alpha,\eta}}
\end{equation}
and
\begin{equation}
\label{eq:covIterative2}
    \bm{\widehat{\phi}_{\beta_i}} = \bm{\Sigma_{\phi_{\beta_i} s_{\alpha}}}\bm{\zeta},
\end{equation}
where $\bm{\zeta}$ in \eref{eq:covIterative1} is computed iteratively. As it stands it requires a MVM per iteration which is prohibitive. Therefore, a fast algorithm for the MVM with these covariance matrices is necessary. In the rest of \S 2, we investigate how it can be suitably mapped into an efficient runtime algorithm.

\subsection{Toeplitz Structure in Covariance Matrices}
An efficient implementation for solving \eref{eq:covIterative1} has been proposed by R. Conan~\cite{Conan-14} for large-scale NGS-based SCAO systems by exploiting the Toeplitz structure of the covariance matrices. Here, we show how these structure is kept in a tomographic setting. 

\begin{figure}
    \centering
    \includegraphics[bb=0 0 354 354, width=0.3\textwidth]{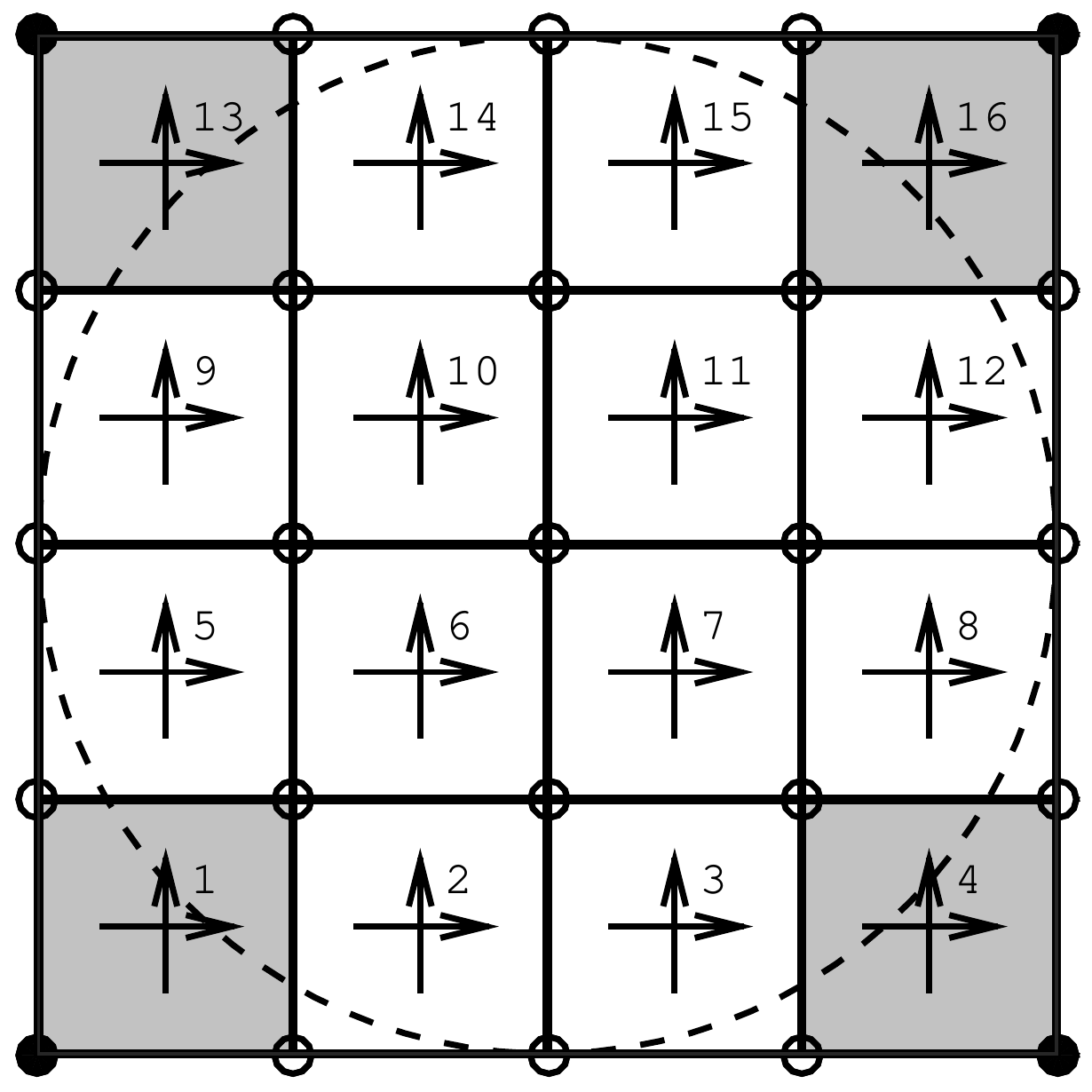}
    \caption{Schematic figure showing a geometry of measurements (arrows) and phase points (circles) for SH-WFS as $n=4$. The dashed lines shows the telescope pupil on the WFS detector. The gray vignetted subapertures and the filled phase points are not taken into account in the WFR.}
    \label{fig:shwfs}
\end{figure}

In order to avoid an overly complicated notation, $\bm{s_{\alpha,\eta}}$, $\bm{s_\alpha}$ and $\bm{s_{\alpha_i}}$ are simplified into $\bm{s_\eta}$, $\bm{s}$ and $\bm{s_i}$, respectively. In addition, we only focus on one target direction $\bm{\beta}$, which is a case of LTAO systems. For multiple target cases such as MOAO systems, WFR for each target direction is performed in parallel. The number of subapertures and phase points in one SH-WFS are $n\times n$ and $(n+1)\times(n+1)$, and the geometric arrangement of the measurement and the phase points shown in \fref{fig:shwfs} as $n=4$. \hlone{First, a square aperture without vignetted subapertures is assumed. The impact of more complicate apertures, such as circular and annular, on the Toeplitz covariance matrix will be discussed in 2.F.}

\subsubsection*{Slope-Slope Covariance Matrix}
\begin{figure}[t]
    \centering
    \includegraphics[bb=0 0 825 360, width=0.49\textwidth]{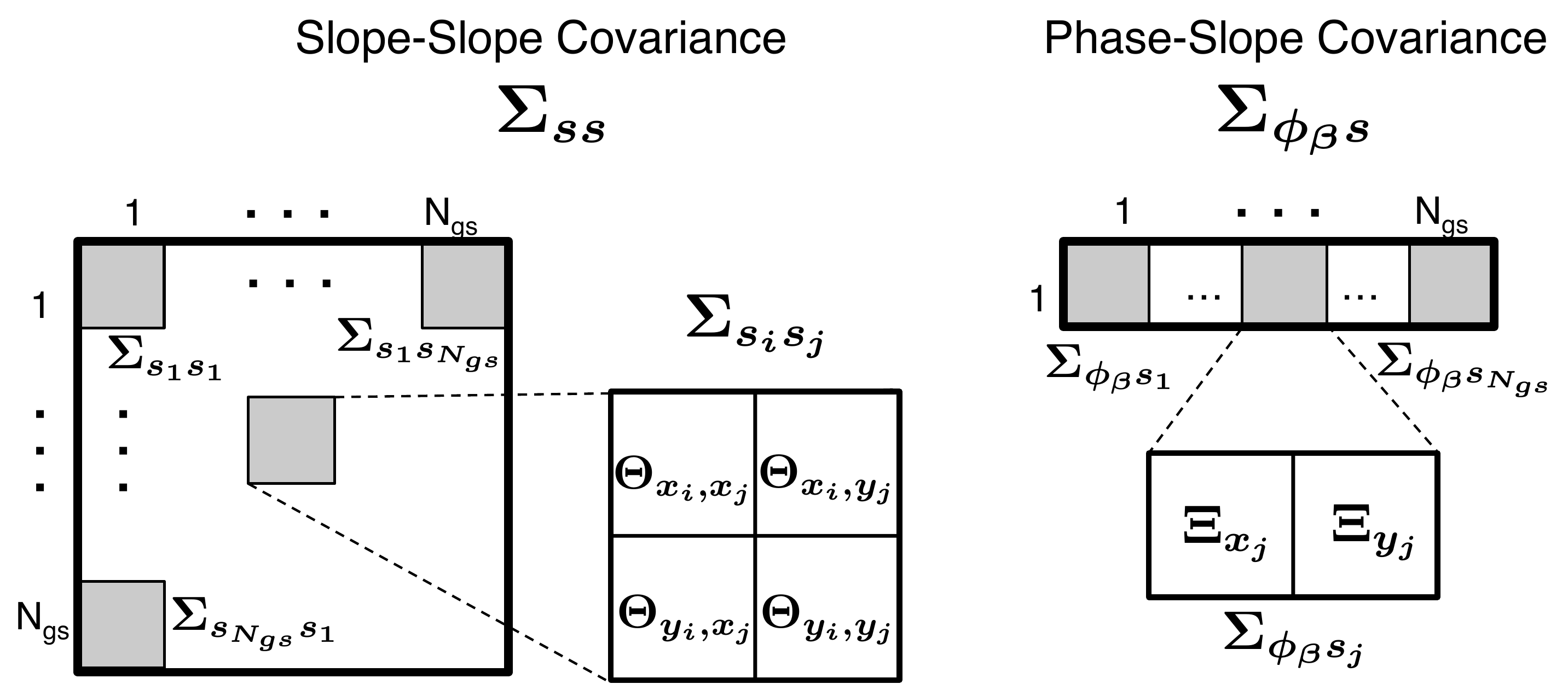}
    \caption{Schematic figure showing the structure of the slope-slope covariance $\bm{\Sigma_{ss}}$ (left) and the phase-slope covariance $\bm{\Sigma_{\phi_{\beta}s}}$ (right).}
    \label{fig:covariance}
\end{figure}

\begin{figure}
    \centering
    \includegraphics[bb=0 0 427 431, width=0.4\textwidth]{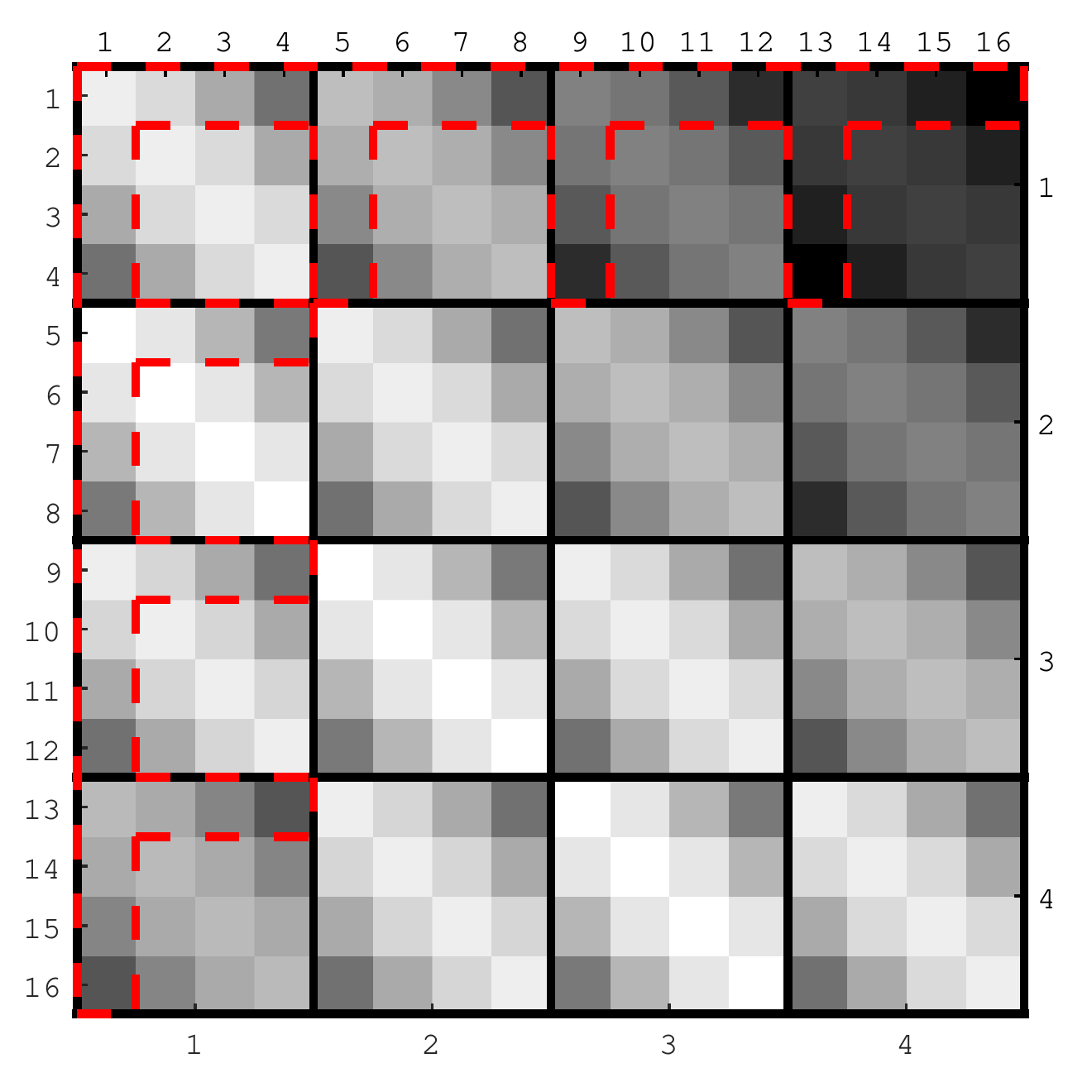}
    \caption{Slope-slope covariance matrix $\bm{\Theta}$ in a case with 4$\times$4 lenslet array. The elements in the red dashed lines are unique elements defining the 2RBT matrix.}
    \label{fig:2rbt_ss}
\end{figure}

As shown in \fref{fig:covariance}, the slope-slope covariance matrix $\bm{\Sigma_{s s}}$ is decomposed into $N_{gs}\times N_{gs}$ blocks, and the $(i,j)$-th block can also be decomposed into $2\times2$ components:
\begin{equation}
    \bm{\Sigma_{s_i s_j}} = 
    \begin{bmatrix}
        \bm{\Theta}_{x_i,x_j} & \bm{\Theta}_{x_i,y_j}\\[5pt]
        \bm{\Theta}_{y_i,x_j} & \bm{\Theta}_{y_i,y_j}
    \end{bmatrix}.
\end{equation}
where $\bm{\Theta_{p,q}}=\langle\bm{s_p}\bm{s_q}^T\rangle$ ($p=x_i,y_j$, $q=x_j,y_j$) with size $n^2\times n^2$.

\begin{figure}[!b]
    \centering
    \includegraphics[bb= 0 0 360 252,width=.42\textwidth]{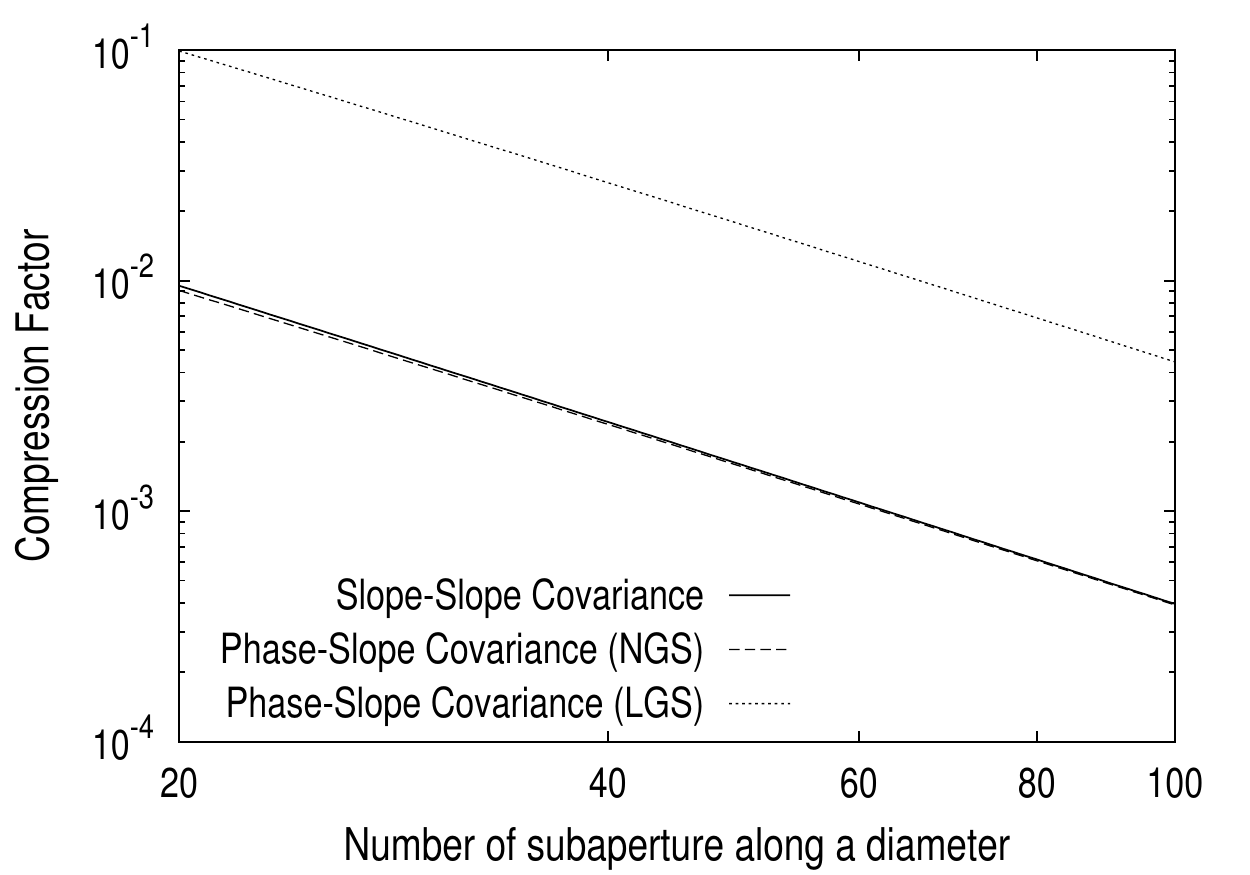}
    \caption{Compression factor (a ratio of the number of unique elements to the total number of elements in the matrix) for $\bm{\Sigma_{s s}}$ (solid line) and $\bm{\Sigma_{\phi_{\beta} s}}$ (dashed line for NGSs case and dotted line for LGSs case). For the LGSs-based, $N_{gs}=6$ and $N_{layer} = 10$ are assumed.}
    \label{fig:num_element}
\end{figure}

Under the assumption of statistically isotropic and homogeneous atmospheric turbulence, a covariance value between two slopes depends only on the spatial separation between the two subapertures. Thanks to this fact, $\bm{\Theta_{p.q}}$ are a two-level Recursive Block Toeplitz (2RBT) matrix, which has $n\times n$ blocks in a Toeplitz arrangement and each block is a Toeplitz matrix with size $n\times n$, as shown in \fref{fig:2rbt_ss} as $n=4$. A Toeplitz matrix with size $n\times n$ are fully defined with $(2n-1)$ elements in its first row and column. Therefore, $\bm{\Theta_{p.q}}$ has $(2n-1)$ unique blocks with $(2n-1)$ unique elements, and hence can be defined only with $(2n-1)^2$ elements (elements surrounded by the red dashed line in \fref{fig:2rbt_ss}), instead of the full elements of $n^4$. As a result, the slope-slope covariance matrix $\bm{\Sigma_{s s}}$, containing $4N_{gs}^2$ of $\bm{\Theta_{p.q}}$, can be compressed into $4N_{gs}^2(2n-1)^2$ elements from $4N_{gs}^2n^4$ full elements. \hlone{The compression factor, which is a ratio of the number of unique elements to the total number of elements in the matrix, of the slope-slope covariance matrix is $(2n-1)^2n^{-4}$} and shown in \fref{fig:num_element} (solid line) as function of $n$. The memory requirements to store the slope-slope covariance matrix is reduced by $6\times10^{-4}$ when $n=80$ by the Toeplitz nature of the covariance matrix.

\hlone{It must be noted that the Toeplitz structure in a covariance matrix $\bm{\Sigma_{xy}}$ holds when the spatial sampling of $x$ and $y$ is the same. The light from a NGS at an infinite altitude propagates through the atmosphere along parallel rays; the spatial sampling of the slope is equal to the subaperture diameter $d$ over all altitudes. On the other hand, the light from a LGS at a finite altitude $h_{lgs}$ propagates spherically along paths creating a cone and hence the spatial sampling changes with altitude $h_k$ by $d_k=d(h_{lgs}-h_k)/h_{lgs}$, i.e. the so-called cone effect. Therefore, a slope-slope covariance matrix between a NGS and a LGS or LGSs at different altitudes (e.g. a sodium LGS and a Rayleigh LGS) is not a 2RBT matrix. This non-Toeplitz structure can also be observed on the phase-slope covariance matrix $\bm{\Sigma_{\phi_{\beta}s}}$ even if we use only LGSs at the same altitude, because the optimization is done for a star located at infinity (parallel rays) but the spatial sampling of $\bm{\phi_\beta}$ is different from $\bm{s}$. This is a key point of this paper and we will discuss how to overcome this limitation for LGSs-based systems in the next section.} \hltwo{We note that in doing so, MCAO systems can also be addressed by out developments.}

\subsubsection*{Phase-Slope Covariance Matrix}

\begin{figure}[t]
    \centering
    \includegraphics[bb=0 0 264 380, width=0.38\textwidth]{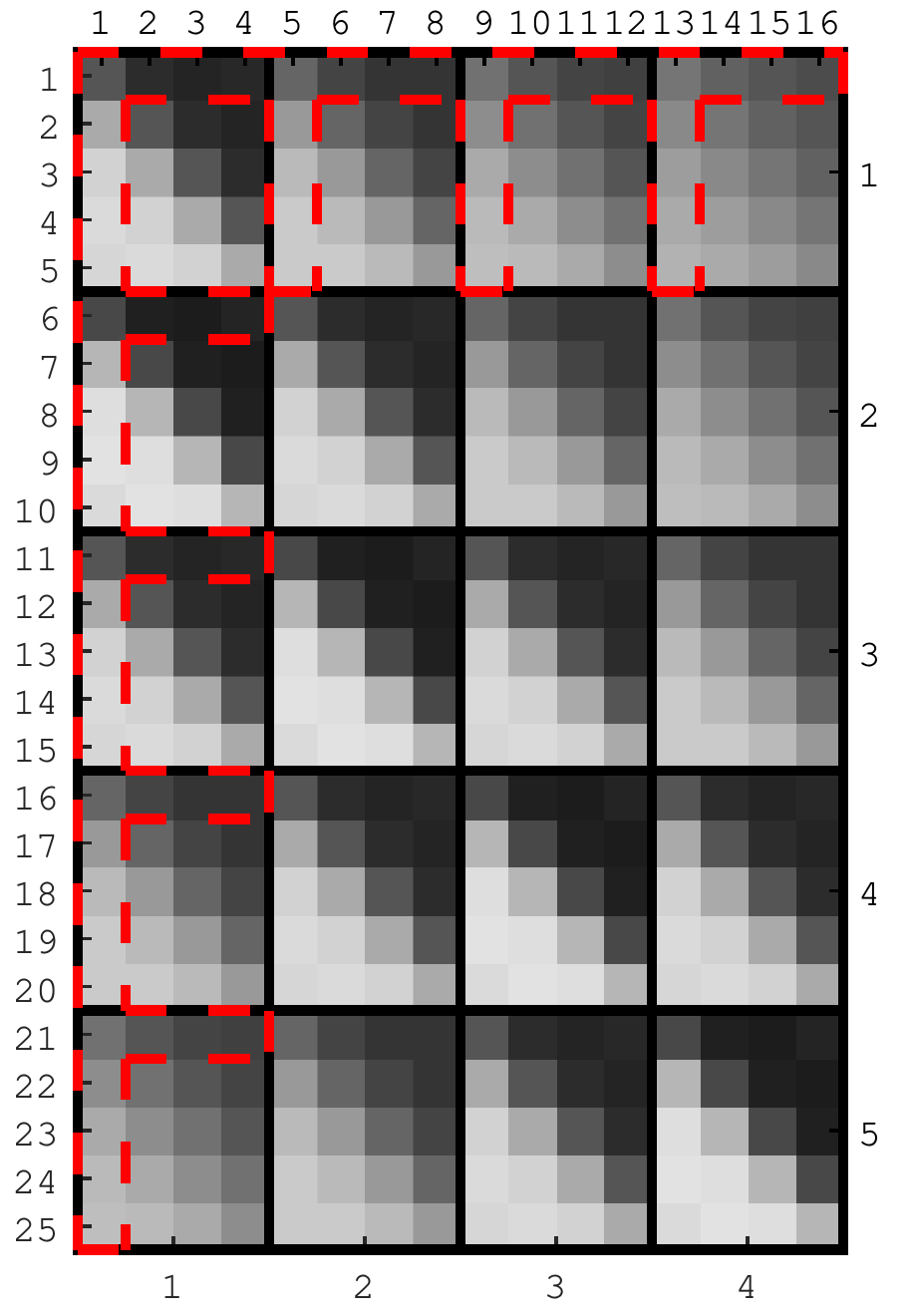}
    \caption{Phase-slope covariance matrix $\bm{\Xi}$ in a case with 4$\times$4 lenslet array. The geometry of the measurements and the reconstructed phase is shown in \fref{fig:shwfs}. The elements in the red dashed lines is unique elements defining the 2RBT matrix.}\label{fig:2rbt_ps}
\end{figure}

As shown in \fref{fig:covariance}, the phase-slope covariance matrix $\bm{\Sigma_{\phi s}}$ has $1\times N_{gs}$ blocks and each block is decomposed into two matrices,
\begin{equation}
    \bm{\Sigma_{\phi s_j}} = 
    \begin{bmatrix}
        \bm{\Xi}_{x_j} & \bm{\Xi}_{y_j}
    \end{bmatrix},
\end{equation}
where $\bm{\Xi_q}=\langle\bm{\phi}\bm{s_q}^T\rangle$ with size $(n+1)^2\times n^2$, following the geometric arrangement shown in \fref{fig:shwfs}. 

For NGSs-based tomographic AO systems, $\bm{\Xi_q}$ is a 2RBT matrice with $(n+1)\times n$ Toeplitz block in $(n+1)\times n$ Toeplitz arrangement. The number of unique elements in $\bm{\Xi_q}$ is $4n^2$ in contrast with $(n+1)^2n^2$ full elements, as shown in \fref{fig:2rbt_ps} as $n=4$ As a result, the phase-slope covariance matrix $\bm{\Sigma_{\phi s}}$ for NGSs-based systems can be defined with $8N_{gs}n^2$ unique elements instead of $2N_{gs}(n+1)^2n^2$ full elements. \hlone{The compression factor for $\bm{\Sigma_{\phi_\beta s}}$ in NGSs-based systems is given by $4(n+1)^{-2}$}, and the memory requirements is reduced by $6\times10^{-4}$ as $n=80$ (\fref{fig:num_element}).

\begin{figure}[t]
    \centering
    \includegraphics[bb=0 0 360 511, width=0.35\textwidth]{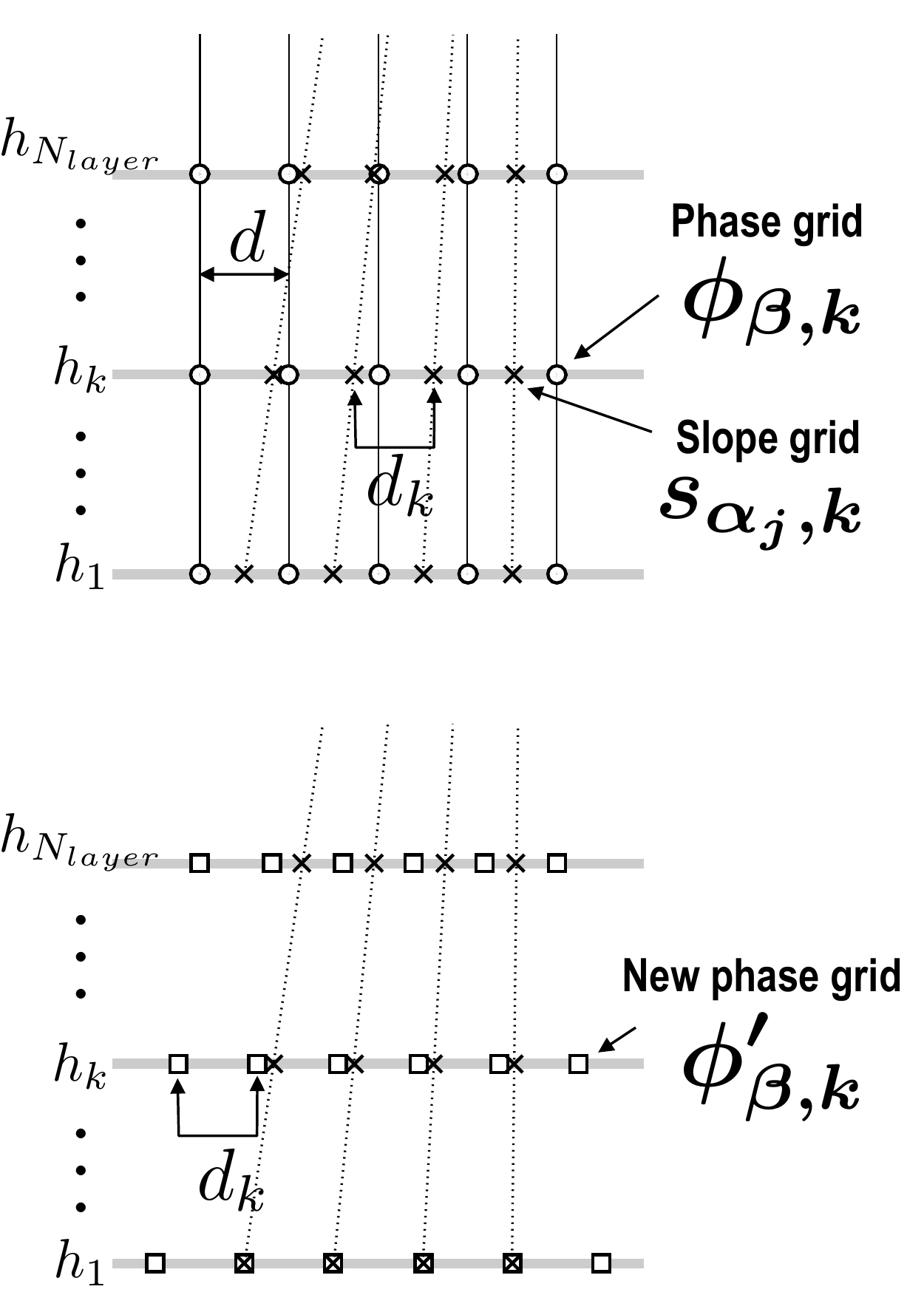}
    \caption{Geometric relation of slope sampling (cross), actual phase sampling (circle) and new phase sampling (square)}\label{fig:grid2}
\end{figure}

As mentioned in the previous section, $\bm{\Xi_q}$ is not Toeplitz matrix for LGSs-based systems due to the spatial sampling variation by the cone effect, as shown in the top panel of \fref{fig:grid2}. In order to overcome this issue, we introduce a new phase vector $\bm{\phi_{\beta,k}'}$ for the $k$-th layer with the same spatial sampling as the LGS slope i.e. $d_k=d(h_{lgs}-h_k)/h_{lgs}$, as shown in the bottom panel of \fref{fig:grid2}. The number of the new phase points is defined $n_k'\times n_k'$ to cover all the original phase points. More points are required at higher altitudes, and the minimum number of $n_k'$ is roughly given by $n_k'=nd/d_k=nh_{lgs}/(h_{lgs}-h_k)$. The new phase vector is connected to the original phase vector through a bi-linear interpolation,
\begin{equation}\label{eq:interp}
    \bm{\phi_{\beta,k}} = \bm{I_k'}\bm{\phi_{\beta,k}'},
\end{equation}
where \hltwo{$\bm{I_k'}$ is a sparse bi-linear interpolation matrix for the new phase grid and has $4n^2$ non-zero elements as well as $\bm{P_{\beta,k}}$ (see 2.B).}

Now we can define new covariance matrices between the new phase vector $\bm{\phi_{\beta,k'}}$ and the slope $\bm{s}$ for the altitude $h_k$ as $\bm{\Xi_q'}=\langle\bm{\phi_{\beta,k}'}\bm{s_q}^T\rangle$, and this covariance matrices is a 2RBT matrix with $(n_k'+n-1)^2$ unique elements. With \eref{eq:interp},
\begin{equation}
    \bm{\Xi_q}=\sum_k^{N_{layer}}\bm{\Xi_{q,k}}=\sum_k^{N_{layer}}\bm{I_k'}\bm{\Xi_{q,k}'}.
\end{equation}


The phase-slope covariance matrix $\bm{\Sigma_{\phi_\beta s}}$ for LGSs-based AO systems can be expressed as a multiplication of $N_{layer}\times 2N_{gs}$ 2RBT matrices $\bm{\Xi_{q,k}'}$ and $N_{layer}$ sparse interpolation matrices $\bm{I'_k}$. The number of unique elements is $\sum_k2N_{gs}(n_k'+n-1)^2$ in 2RBT matrices, where $\sum_k$ is an integration over $N_{layer}$ altitudes, and $4N_{layer}n^2$ in the sparse interpolation matrices. \hlone{The compression factor of $\bm{\Sigma_{\phi_\beta s}}$ in LGSs-based systems is given by $[\sum_k(n_k'+n-1)^2](n+1)^{-2}n^{-2}+2N_{layer}N_{gs}^{-1}(n+1)^{-2}$}. If we assume $N_{gs}=6$ and $N_{layer} = 9$, the compression factor is $6\times10^{-3}$ as $n=80$ (\fref{fig:num_element}).

\hltwo{If we extend the new phase grid to cover the range of the layered phase and remove the bi-linear interpolation by $\bm{I_k'}$, the layered phase at each altitude can be estimated by the spatio-angular reconstructor with the 2RBT formulation. This needs more computation than the directional phase reconstructor, but can be applicable to MCAO systems.}

\subsection{Efficient MVM with 2RBT Matrix}
\label{sec:mvm_2rtb}

The algorithm to compute efficiently the MVM when the matrix is
2RBT was originally developed in \cite{Lee-86}, and later applied to WFR in \cite{Conan-14}. In this section, we briefly summarize this algorithm and show how to introduce this algorithm into our tomographic WFR.

Let us consider a product of a $m^2\times n^2$ 2RBT matrix $\bm{T}$
and a vector $\bm{x}$ of length $n^2$ i.e. $\bm{T}\bm{x}=\bm{y}$,
where $\bm{y}$ is a vector of length $m^2$. The Toeplitz matrix $\bm{T}$ has
$m\times n$ blocks in a Toeplitz arrangement and each block is a
$m\times n$ Toeplitz matrix, and therefore, can be
replaced by a vector $\bm{t}$ containing $(n+m-1)^2$ unique elements
of $\bm{T}$. The MVM $\bm{T}\bm{x}=\bm{y}$ is
obtained through FFT $\bm{\widetilde{t}}$ of $\bm{t}$,
conveniently denoted by a tilde symbol.

The MVM $\bm{T}\bm{x}=\bm{y}$ is accomplished
by the steps following:
\begin{enumerate}
    \item a vector $\bm{x}$ is shuffled into $\bm{b}$ of length $(m+n-1)^2$ according to the rules:
    \begin{align}\label{eq:2rbtmvm1}
        \bm{b}_{\mu(i,j)}&=\bm{x}_{k(i,j)}\ \ \ \ (0\geq i,j\geq n),\\
        k(i,j)&=in+j,\\
        \mu(i,j)&=(m+n)(n-1)-i(m+n-1)-j,
    \end{align}
    where elements of $\bm{b}$ except for $\mu(i,j)$ are zero. Defining $\bm{S_1}$ as a shuffling operator converting $\bm{x}$ to $\bm{b}$, we can rewrite the step 1 as $\bm{b}=\bm{S_1x}$. The size of $\bm{S_1}$ is $(m+n-1)^2\times n^2$ and elements indexed with $[\mu(i,j),k(i,j)]$ are 1, otherwise 0, and therefore $\bm{S_1}$ is a sparse matrix with $n^2$ non-zeros elements.
    \item the 1-D FFT of $\bm{b}$ is computed
    \begin{equation}
        \bm{\widetilde{b}}=\mathcal{F}[\bm{b}]
    \end{equation}
    \item the element-wise vector product (represented by $\cdot$) of $\bm{\widetilde{t}}$ and $\bm{\widetilde{b}}$ is computed
    \begin{equation}
        \bm{\widetilde{c}}=\bm{\widetilde{t}}\cdot\bm{\widetilde{b}}
    \end{equation}
    \item the inverse 1-D FFT of $\bm{c}$ is computed
    \begin{equation}
        \bm{c}=\mathcal{F}^{-1}[\bm{\widetilde{c}}]
    \end{equation}
    \item $\bm{c}$ is reshuffled into $\bm{y}$ according to the rules:
    \begin{align}
        &\bm{y}_{k(i,j)}=\bm{c}_{\mu(i,j)}\ \ \ \ (0\geq i,j\geq m),\\
        &k(i,j)=im+j,\\
        &\begin{aligned}
            \mu(i,j)&=(m+n)(m+n-1)\\
            &-(i+1)(m+n-1)-(j+1).
        \end{aligned}
    \end{align}
    In the same way as the step 1, the step 5 can be expressed as $\bm{y}=\bm{S_2}\bm{c}$ with a sparse reshuffling $\bm{S_2}$ of size $m^2\times(m+n-1)^2$. The number of non-zeros elements in $\bm{S_2}$ is $m^2$. 
\end{enumerate}

Finally, the computation flow from the step 1 to 5 can be summarized as follows
\begin{align}\label{eq:2rbtmvm}
    \bm{y}&=\bm{Tx}=\bm{S_2}\mathcal{F}^{-1}\left[\bm{\widetilde{t}}\cdot\bm{\widetilde{b}}\right]\\
    \bm{\widetilde{b}}&=\mathcal{F}[\bm{S_1}\bm{x}].
\end{align}

The 2RBT MVM consists of two 1-D FFTs (forward and backward), one element-wise vector product and two shufflings (i.e. two sparse MVM), and hence the number of operation in the 2RBT MVM is $(m+n-1)^2[4a\log(m+n-1)+1]+m^2+n^2$ instead of $(m+n-1)^4$ of full MVM, assuming that the number of operation in the FFT is $aN\log N$ with $N$ is the data size, and that the number of operation in one sparse MVM is equal to the number of non-zeros elements. The coefficient $a$ depends on the FFT algorithm choice and has a value of ranging between 4 and 5 typically \cite{soni11}.

\subsubsection*{2RBT MVM for Slope-Slope Covariance Matrix}
We now apply the efficient 2RBT MVM algorithm to the tomographic computation. With respect to $(\bm{\Sigma_{ss}}+\bm{\Sigma_{\eta\eta}})\bm{\zeta}$ in \eref{eq:covIterative1}, we focus on $\bm{\Sigma_{ss}}\bm{\zeta}$ because in general $\bm{\Sigma_{\eta\eta}}$ is a very sparse matrix due to the assumption that measurement noise is both temporally and spatially statistically independent with the computational cost of $\bm{\Sigma_{\eta\eta}}\bm{\zeta}$ being less of an issue. The product of $\bm{\Sigma_{ss}}\bm{\zeta}$ is a concatenation of 
\begin{align}\label{eq:mvm_toep1}
    \bm{\Sigma_{s_is}}\bm{\zeta}=
    \begin{bmatrix}
        \sum_j^{N_{gs}}(\bm{\Theta_{x_i,x_j}}\bm{\zeta_{x_j}}+\bm{\Theta_{x_i,y_j}}\bm{\zeta_{y_j}})\\
        \sum_j^{N_{gs}}(\bm{\Theta_{y_i,x_j}}\bm{\zeta_{x_j}}+\bm{\Theta_{y_i,y_j}}\bm{\zeta_{y_j}})
    \end{bmatrix},
\end{align}
where $i=(1,\cdots,N_{gs})$. In \eref{eq:mvm_toep1} each $\bm{\Theta_{p,q}}\bm{\zeta_{q}}$ is a 2RBT MVM. Defining $\bm{\theta_{p,q}}$ as the unique element vector of $\bm{\Theta_{p,q}}$, $\bm{\widetilde{\theta}_{p,q}}$ as its FFT and $\bm{\widetilde{\zeta}_q} = \mathcal{F}[\bm{S_1}\bm{\zeta_q}]$, we can rewrite \eref{eq:mvm_toep1} with the 2RBT MVM formulation as
\begin{align}\label{eq:mvm_ss}
    &\bm{\Sigma_{s_is}}\bm{\zeta}=
    \begin{bmatrix}
        \bm{S_2}\mathcal{F}^{-1}\left\{\sum_j^{N_{gs}}(\bm{\widetilde{\theta}_{x_i,x_j}}\cdot\bm{\widetilde{\zeta}_{x_j}}+\bm{\widetilde{\theta}_{x_i,y_j}}\cdot\bm{\widetilde{\zeta}_{y_j}})\right\}\\
        \bm{S_2}\mathcal{F}^{-1}\left\{\sum_j^{N_{gs}}(\bm{\widetilde{\theta}_{y_i,x_j}}\cdot\bm{\widetilde{\zeta}_{x_j}}+\bm{\widetilde{\theta}_{y_i,y_j}}\cdot\bm{\widetilde{\zeta}_{y_j}})\right\}\\
    \end{bmatrix}.
\end{align}

In an actual computation, $\bm{\widetilde{\zeta}_q} = \mathcal{F}[\bm{S_1}\bm{\zeta_q}]$ is computed first for all $q$ and reused repeatedly in the following steps to avoid redundant computations. Moreover, thanks to the linearity of the FFT, the inverse FFT and the reshuffling in the step 4 and 5 of the algorithm can be performed in the end after the summation for $j$, as already shown in \eref{eq:mvm_ss}, to reduce the number of the FFT. As a result, the MVM of $\bm{\Sigma_{ss}}\bm{\zeta}$ consists of $2N_{gs}$ forward and backward 1-D FFTs with length $(2n-1)^2$, $4N_{gs}^2$ element-wise vector product with length $(2n-1)^2$, $2N_{gs}$ shuffling and reshuffling, and the total required number of operations in $\bm{\Sigma_{ss}}\bm{\zeta}$ is $4N_{gs}\{(2n-1)^2[2a\log(2n-1)+N_{gs}]+n^2\}$ instead of $4N_{gs}^2n^4$ of the full MVM. The reduction factor in the number of operations, which is a ratio of the number of operation for the 2RBT MVM to one for the full MVM, for $\bm{\Sigma_{ss}}\bm{\zeta}$ is $\{(2n-1)^2[2a\log(2n-1)+N_{gs}]+n^2\}N_{gs}^{-1}n^{-4}$, shown in \fref{fig:num_op} (solid line), and $8\times10^{-3}$ as $n=80$ and $a=5$.

\subsubsection*{2RBT MVM for Phase-Slope Covariance Matrix}
\begin{figure}
    \centering
    \includegraphics[bb= 0 0 360 252, width=.46\textwidth]{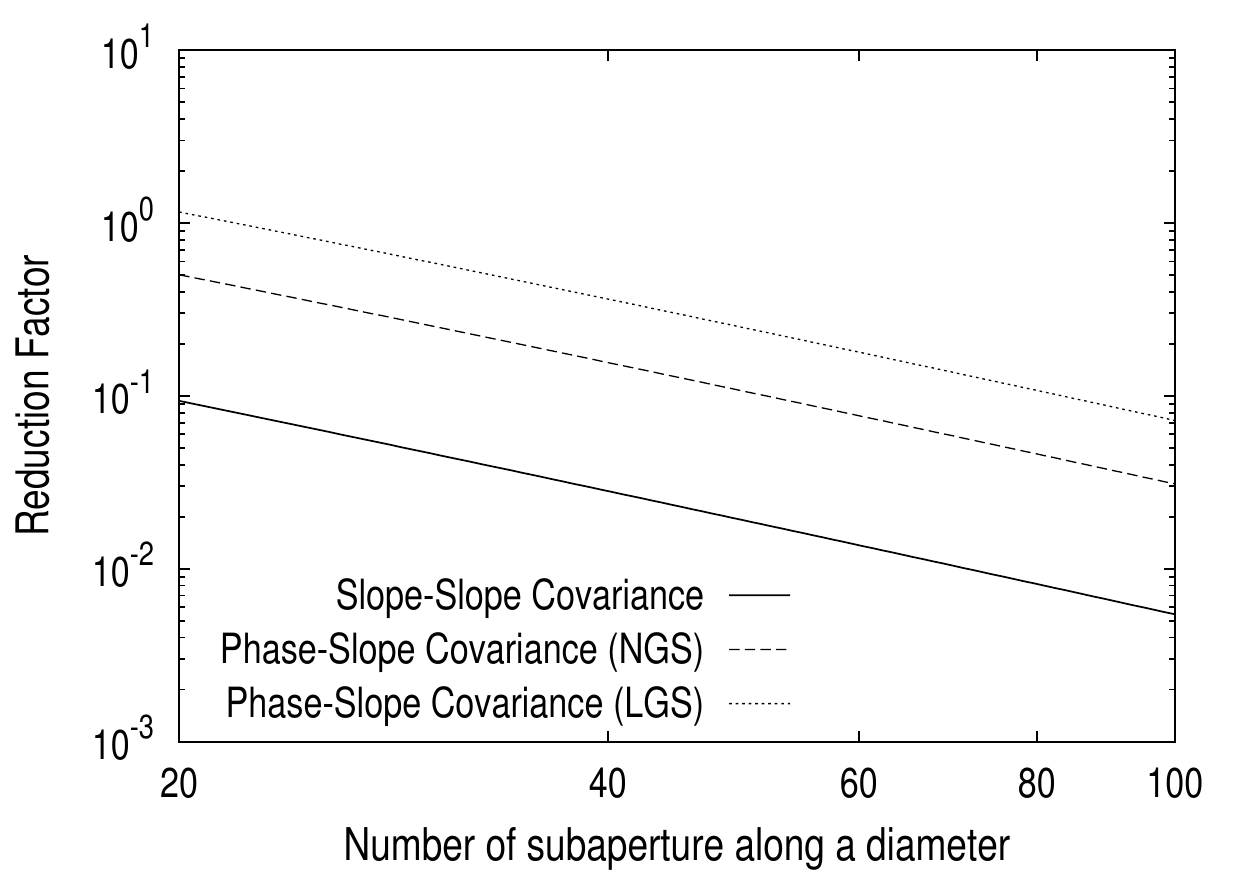}
    \caption{Reduction factor of the number of operations required for $\bm{\Sigma_{s s}}\bm{\zeta}$ (solid line) and $\bm{\Sigma_{\phi s}}\bm{\zeta}$ (the dashed line for NGSs case and the dotted line for LGSs case) as function of the number of subapertures across telescope diameter, assuming $N_{gs}=6$, $N_{layer}=9$ and $n'=1.2n$ ($h_{lgs}=90$\,km and $h_{N_{layer}}=16$\,km). The coefficient for the FFT is assumed $a=5$. }\label{fig:num_op}
\end{figure}

In a NGSs-based case, $\bm{\Sigma_{\phi_{\beta} s}}\bm{\zeta}$ in \eref{eq:covIterative2} can be formulated to the 2RBT MVM in the same way as $\bm{\Sigma_{s s}}\bm{\zeta}$,
\begin{align}\label{eq:mvm_ps}
    \bm{\Sigma_{\phi_\beta s}}\bm{\zeta}&=
        \bm{S_2}\mathcal{F}^{-1}\left\{\sum_j^{N_{gs}}(\bm{\widetilde{\xi}_{x_j}}\cdot\bm{\widetilde{\zeta}_{x_j}}+\bm{\widetilde{\xi}_{y_j}}\cdot\bm{\widetilde{\zeta}_{y_j}})\right\},
\end{align}
where $\bm{\xi}_q$ is a unique elements vector of $\bm{\Xi_q}$ and $\bm{\widetilde{\zeta}_q}=\mathcal{F}[\bm{S_1}\bm{\zeta_q}]$. This formulation contains $2N_{gs}$ forward FFTs and 1 backward FFT with length
 $4n^2$, $2N_{gs}$ element-wise vector product with length $4n^2$, $2N_{gs}$ shuffling and 1 reshuffling, and the total required number of operations is $2N_{gs}[4n^2(2a\log2n+1)+n^2]+8n^2\log2n+(n+1)^2$ instead of $2N_{gs}(n+1)^2n^2$ of the full MVM. The reduce factor is shown by the dashed line in \fref{fig:num_op} and $4\times10^{-2}$ as $n=80$ and $a=5$.

In LGSs-based AO systems, the number of phase points on the new grid $n_k'$ is different depending on altitudes and, therefore the shuffling and reshuffling operator $S_1$ and $S_2$ should be defined for each altitude like $S_{1,k}$ and $S_{2,k}$. This means that $\bm{\Sigma_{\phi_\beta s}}\bm{\zeta}$ requires to compute 2RBT MVMs separately for each altitude followed by the interpolation. The 2RBT MVM formulation for $\bm{\Sigma_{\phi_\beta s}}\bm{\zeta}$ with LGSs is
\begin{align}\label{eq:ToeplitzLGS}
    \bm{\Sigma_{\phi s}}\bm{\zeta}&=\sum_k^{N_{layer}}
        \bm{I_k'}\bm{S_{2,k}}\mathcal{F}^{-1}\left\{\sum_j^{N_{gs}}(\bm{\widetilde{\xi'}_{x_j,k}}\cdot\bm{\widetilde{\zeta}_{x_j,k}}+\bm{\widetilde{\xi'}_{y_j,k}}\cdot\bm{\widetilde{\zeta}_{y_j,k}})\right\}.
\end{align}
The forward FFT $\bm{\widetilde{\zeta}_{q,k}}=\mathcal{F}[\bm{S_{1,k}}\bm{\zeta_{q,k}}]$ should also be computed for each altitude because of the different $n_k'$ depending on the altitudes unlike $\bm{\Sigma_{ss}}$ and $\bm{\Sigma_{\phi_\beta s}}$ for a NGSs case, and it causes a huge computational complexity even more than the complexity of the full MVM.

In order to avoid this issue, all $n_{\phi,k}'$ should be set to the same value $n'$, where $n'=n_{\phi,N_{layer}}'$ ($n_k'$ at the highest altitude $h_{N_{layer}}$ ) to cover the all original phase points at all altitudes. This causes extra phase points for lower altitude layers (see \fref{fig:grid2}) but we can use the same shuffling and reshuffling operator for all altitudes and $\bm{\widetilde{\zeta}_{q,k}}=\mathcal{F}[\bm{S_1}\bm{\zeta_{q,k}}]$ is needed to be computed just once. The number of non-zero values in $\bm{I_k'}$ and $\bm{I_k'S_2}$ are the same because $\bm{S_2}$ just shuffle the order of rows of $\bm{I_k'}$, and hence we can precompute $\bm{I_k'S_2}$ offline. Then, the required computations for one $\bm{\Sigma_{\phi_\beta s}}\bm{\zeta}$ in LGSs-based system with the 2RBT formulation are $2N_{gs}$ forward FFTs with length $(n'+n-1)^2$, $2N_{gs}\times N_{layer}$ element-wise vector products with length $(n'+n-1)^2$, $N_{layer}$ backward FFT with length $(n'+n-1)^2$, $2N_{gs}$ shuffling and $N_{layer}$ interpolation (including reshuffling). The total number of operation is $2(n'+n-1)^2[(2N_{gs}+N_{layer})a\log(n'+n-1)+N_{gs}N_{layer}]+2N_{gs}n^2+4N_{layer}(n+1)^2$. The reduce factor is shown by the dotted line in \fref{fig:num_op} and 0.1 as $n=80$, $a=5$ and $n'=1.2n$ (assuming $h_{lgs}=90$\,km and $h_{N_{layer}}=16$\,km).




\subsection{Non-square aperture, mode removal and noise-weighted reconstruction for elongated LGS spots}\label{sec:2f}

\subsubsection*{Non-square aperture}
Although the discussion above assumes a square aperture, apertures have generally more complicated shapes such as circular and annular thus vignetting a portion of subapertures and phase points (shown in \fref{fig:shwfs}). These need to be removed from the tomographic WFR.

The number of valid subapertures and phase points at the aperture-plane are denoted by $n_{vs}$ and $n_{v\phi}$, respectively. To take into account the effect of the non-square aperture in our 2RBT formulation, we introduce a sparse aperture masking matrix $\bm{W_{s,i}}$ with size $2n_{vs}\times n^2$ for $i$-th WFS measurements. Only $2n_{vs}$ elements in $\bm{W_{s,i}}$ are 1 and otherwise 0 to extract valid measurements (in $x$ and $y$ directions) from vectors. With $\bm{W_{s,i}}$, we redefine the vignetted noisy measurement as
\begin{equation}
    \bm{s_{\alpha_i,\eta}} = \bm{W_{s,i}}(\bm{s_{\alpha_i}}+\bm{\eta}).
\end{equation}
The noisy measurements $\bm{s_{\alpha_i,\eta}}$ contains only $n_{vs}$ valid measurements, wheres $\bm{s_{\alpha_i}}$ and $\bm{\eta}$ are defined with a full square aperture. Denoting $\bm{W_s}$ as the diagonal block matrix of all $\bm{W_{s,i}}$ and $\bm{W_\phi}$ as an aperture matrix for phase points, we can write the spatio-angular reconstruction in \eref{eq:covIterative1} and \eref{eq:covIterative2} as
\begin{align}
    \label{eq:SAap1}
    &\bm{W_s}\left(\bm{\Sigma_{s_{\alpha}s_{\alpha}}}+\bm{\Sigma_{\eta\eta}}\right)\bm{\zeta_W}=\bm{s_{\alpha_i,\eta}}\\
    \label{eq:SAap2}
    &\bm{\widehat{\phi}_{\beta_i}} = \bm{W_\phi}\bm{\Sigma_{\phi_{\beta_i} s_{\alpha_i}}}\bm{\zeta_W},
\end{align}
where $\bm{\zeta}$ is replaced by $\bm{\zeta_W}=\bm{W_s}\bm{\zeta}$.

Although this formulation requires additional computations involving $\bm{W_s}$ and $\bm{W_\phi}$, the vignetted measurements are totally removed from the tomographic WFR whilst keeping the Toeplitz structure of the covariance matrices. For \eref{eq:SAap1}, $\bm{\Sigma_{ss}}\bm{\zeta}$ is replaced by $\bm{W_s}\bm{\Sigma_{ss}}\bm{\zeta_W}$, and the required number of additional operations is $2N_{gs}n_{vs}$, which is less than 1\,\% of that in a 2RBT MVM of $\bm{\Sigma_{ss}}\bm{\zeta_W}$. For \eref{eq:SAap2}, $\bm{\Sigma_{\phi_\beta s}}\bm{\zeta}$ is replaced by $\bm{W_\phi}\bm{\Sigma_{\phi_\beta s}}\bm{\zeta_W}$, and only $n_{v\phi}$ additional operations are needed for the non-square aperture. Therefore, this aperture operation doesn't affect the total computational complexity.

\subsubsection*{Mode removal}
When using LGSs, tip, tilt and focus modes are not measurable and are supplemented by low-order NGS(s) measurements. Under a split tomography control approach \cite{gilles08a}, these modes should be removed from the measurement and consequently from the reconstruction.

Let us introduce $\bm{Z_{s,i}}$ as a tip/tilt/focus modal matrix for $i$-th WFS. Each column of $\bm{Z_{s,i}}$ contains a vectorized mode defined with the non-vignetted valid subapertures. The size of $\bm{Z_{s,i}}$ are $2n_{vs}\times 3$. The tip/tilt/focus projection matrix is given by $\bm{Z_{s_i}}\bm{Z_{s_i}}^\dag$ \cite{ellerbroek02}, \hlone{where $^\dag$ means pseudo inverse}. Then, the mode removal matrix $\bm{M_{s,i}}$ for $i$-th WFS measurement is
\begin{equation}
    \bm{M_{s,i}}=\bm{I}-\bm{Z_{s_i}}\bm{Z_{s_i}}^\dag,
\end{equation}
where $I$ is the identity matrix of appropriate size. The block diagonal matrix containing all $\bm{M_{s,i}}$ is denoted $\bm{M_{s}}$ and the removal matrix in phase space is denoted by $\bm{M_\phi}$ with size $n_{v\phi}\times 3$.

The mode removal is actually performed in the same way as the aperture masking by replacing $\bm{W_s}$, $\bm{W_\phi}$ and $\bm{\zeta_W}$ in \eref{eq:SAap1} and \eref{eq:SAap2} with $\bm{M_s}\bm{W_s}$, $\bm{M_\phi}\bm{W_\phi}$ and $\bm{\zeta_{MW}}=\bm{M_s}\bm{\zeta_W}$. In $\bm{M_s}\bm{W_s}\bm{\Sigma_{ss}}\bm{\zeta_W}$, the additional computations due to the mode removal are $N_{gs}$ of two MVMs with $\bm{Z_{s_i}}$ and $\bm{Z_{s_i}}^\dag$ in $\bm{M_{s,i}}$. The modal matrices and its pseudo inverse are not sparse yet very low-rank, and therefore the mode removal should not affect the total computational complexity. The required number of additional operation due to the mode removal are $6N_{gs}n_{vs}$ for the slope-slope covariance MVM and $3n_{v\phi}$ for the phase slope covariance MVM.

\subsubsection*{Noise-weighted reconstruction for elongated LGS spots}

The elongation of LGS spots on SH-WFS causes additional differential noise depending on the subaperture positions with respect to the laser launch telescope and a cross-correlation term between x- and y-slope in every subaperture. This should be taken into account in the noise covariance matrix. The latter is a diagonal matrix in the NGS case and turns into a 2$\times$2 block diagonal matrix in the LGS case, which is still a sparse matrix. This difference in the noise covariance doesn't affect the Topelitz nature of the other covariance matrices.

\subsection{Iterative algorithm}

We have used standard iterative algorithms for the class of problems we are dealing with \cite{saad03, vogel02}. We tested in particular the Conjugate Gradient (CG), minimum residual method (MINRES) and bi-conjugate gradient stabilized method (BICGSTAB). All these methods were run without pre-conditioning, although suitable formulations could be thought of provided the structure of the operations involved \cite{ellerbroek09a, vogel04a}.

\begingroup
\renewcommand{\arraystretch}{1.1}
\begin{table}[b]
    \centering
    \begin{tabular}{|l|l|}
    \hline
        \textbf{Telescope}   & Aperture Diameter: $D=8$\,m \\
                    & Central Obstruction: 30\% \\
                    & Zenith angle: $z=0$\,degree \\
    \hline
        \textbf{Guide stars} & 3 NGSs or LGSs at 90\,km \\
                    & Asterism radius: 10--50\,arcsec\\
    \hline
        \textbf{WFS}         & 3 Shack-Hartmanns \\
                    & $16\times16$ subapertures/WFS \\
    \hline
        \textbf{Atmosphere}  & Fried parameter: $r_0=0.15$\,m \\
                    & Outer scale: $\mathcal{L}_0=25$\,m \\
                    & Altitudes: [0, 5, 10]\,km \\
                    & $C_N^2$ fraction [0.5 0.25 0.25] \\
    \hline
        \textbf{Control}     & Open-loop \\
                    & no temporal-delay \\
    \hline
    \end{tabular}
    \caption{Parameters for analytical computation and simulation used in Section 3.}
    \label{tab:simu_param1}
\end{table}
\endgroup

\section{Performance comparison on a  8\,m telescope AO system}\label{sec:perfComparison}

\begin{figure*}[t]
	\centering
	\includegraphics[bb=0 0 360 252,width=.43\textwidth]{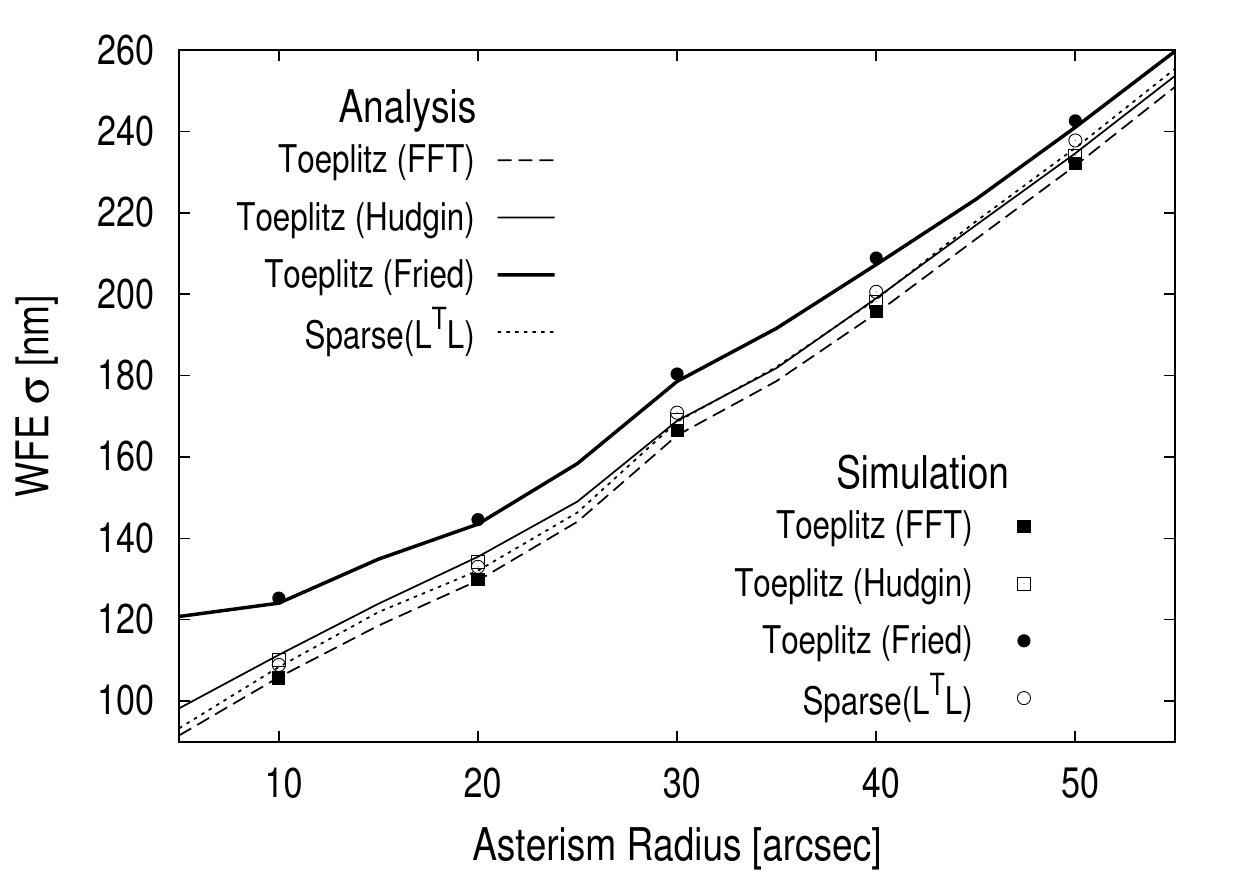}
	\hspace{0.03\textwidth}
	\includegraphics[bb=0 0 360 252,width=.43\textwidth]{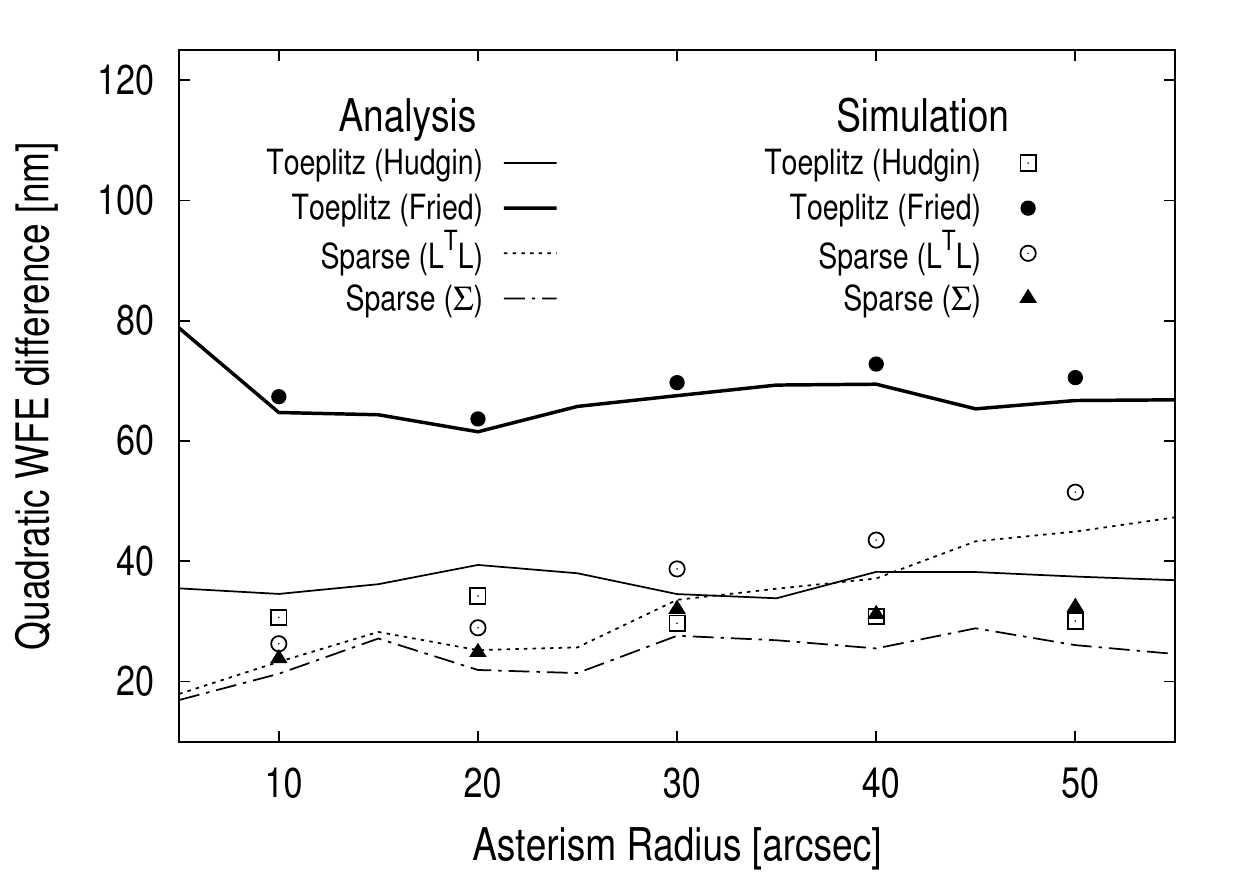}
	\caption{Left: WFE with different algorithm as function of asterism of guide stars. The lines shows the analytical results and the points shows the results from numerical simulation. Right: Quadratic WFE difference from the WFE of the Toeplitz method with the FFT gradient model. The symbols are the same as those in the left panel, except for the dot-dashed line and triangles showing the result with the sparse method with non-approximated real regularization term.}
	\label{fig:performance1}
	\vspace{5mm}
	\includegraphics[bb=0 0 360 252,width=.43\textwidth]{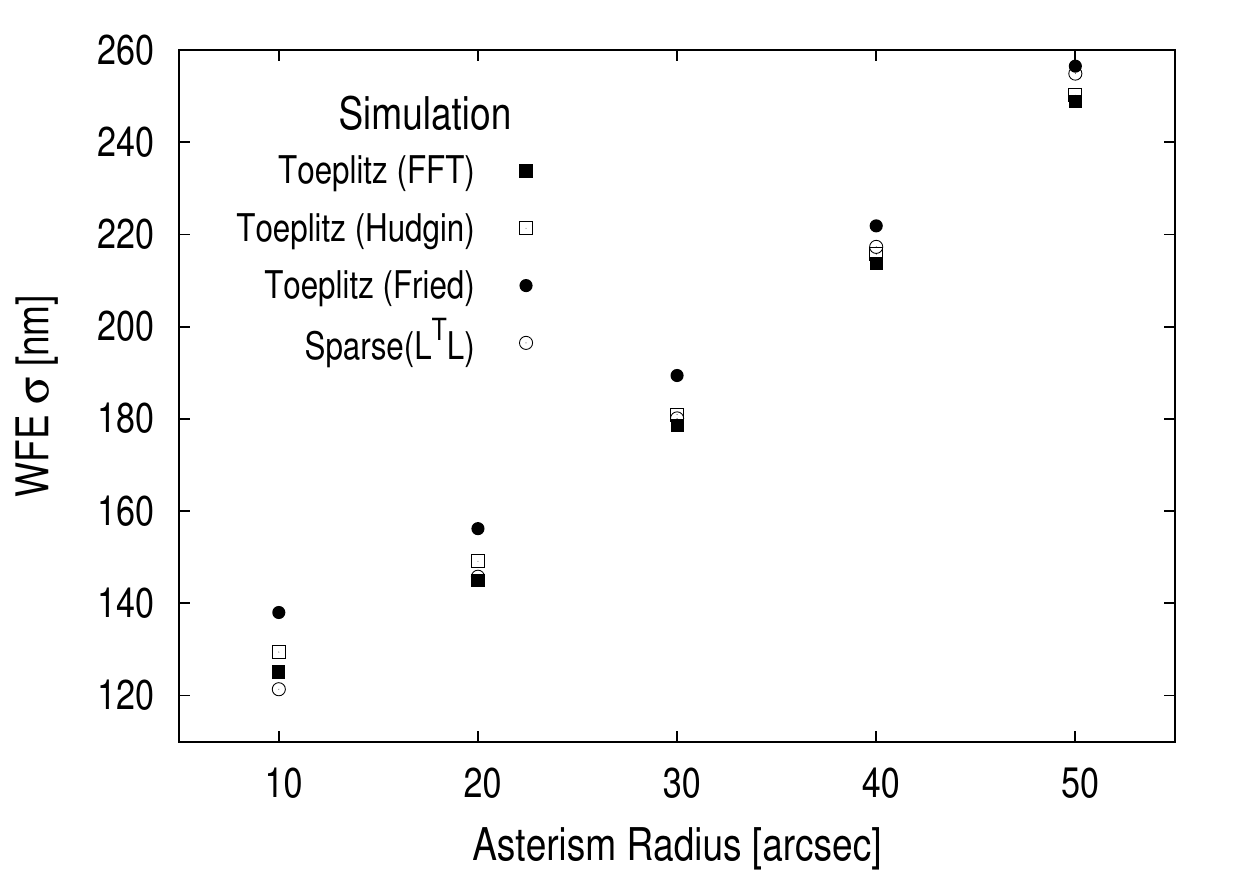}
	\hspace{0.03\textwidth}
	\includegraphics[bb=0 0 360 252,width=.43\textwidth]{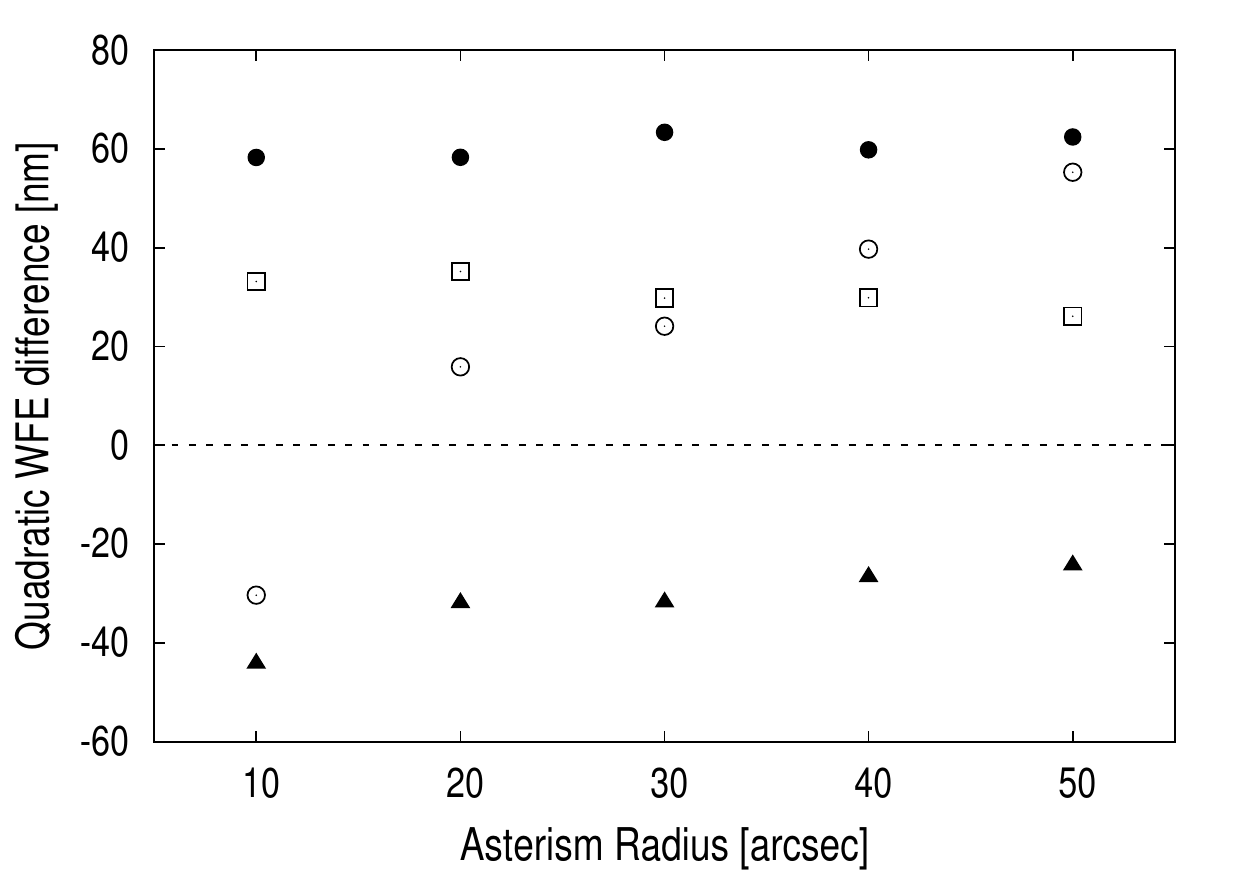}
	\caption{WFE (left) and quadratic WFE difference from the WFE of the Toeplitz method with the FFT gradient model (right) for the LGS case. Only the result from the simulation is plotted. The symbols are the same as in \fref{fig:performance1}.}
	\label{fig:performance2}
\end{figure*}

As a first comparative illustrative case, we compare the performance of the Toeplitz-based reconstructor to a sparse-based reconstructor
using both Monte Carlo simulations obtained with the Object-Oriented Matlab AO simulator (OOMAO) \cite{Conan14} to the results from the analytic expression for the statistical wave-front error (WFE)
\begin{align}
    \label{eq:analytical1}
    \sigma^2_\beta&=\left\langle\left|\left|\bm{\phi}_{\beta}-\bm{\widehat{\phi}}_{\beta}\right|\right|^2_{L_2(\Omega)}\right\rangle\\
    &=\text{Trace}\left(\bm{\Sigma_{\phi_\beta\phi_\beta}}-\bm{R_\beta}\bm{\Sigma_{s_\alpha \phi_\beta}}-\bm{\Sigma_{\phi_\beta s_\alpha}}\bm{R_\beta}^T+\bm{R_\beta}\bm{\Sigma_{s_\alpha s_\alpha}}\bm{R_\beta}^T\right)\notag\\
    \label{eq:analytical2}
    &\hspace{5mm}+\text{Trace}\left(\bm{R}\bm{\Sigma_{\eta\eta}}\bm{R}^T\right)\\
    \label{eq:analytical3}
    &=\sigma^2_{\beta, tomo}+\sigma^2_{\beta, noise}
\end{align}
where $\sigma^2_{\beta, tomo}$ is the tomographic WFE depending on an asterism of GSs and the turbulence model assumed in the reconstructor and $\sigma^2_{\beta, noise}$ is the noise propagation through the reconstructor.

Here, we assume a tomographic system with parameters given in Table \ref{tab:simu_param1}; the projection onto DM, being a common step to all methods, is not taken into account at this time. Three guide stars are placed on a triangle with varying radius (separation from the center), and the WFE is evaluated only for the on-axis direction, $\bm{\beta}=(0,0)$. \hltwo{For a LGSs-case, the low-order modes removal is taken into account.}

For the Toeplitz-based methods we investigate the impact of the gradient model: we compare an accurate, non-sparse model whose definition is more conveniently done in the spatial-frequency domain (therefore the FFT model) to sparse discrete approximations further developed in Appendix A. The gradient model in the analytic expression (i.e. $\bm{\Sigma_{s\phi}}$, $\bm{\Sigma_{s\phi}}$ and $\bm{\Sigma_{\phi s}}$ in \eref{eq:analytical2}) is assumed to be the FFT model which is the most accurate; only in the reconstructor this model is subject to change.

The left panel of \fref{fig:performance1} shows $\sigma^2$ for a NGSs-based tomographic system as function of the asterism radius. The lines in \fref{fig:performance1} presents analytic WFE from \eref{eq:analytical1} and the points show the WFE from the numerical simulation. First, the analytic lines match well the simulation results, indicating that the FFT-gradient model is accurate enough for a reference slope in a NGS-based case. As for the gradient models of the Toeplitz-based method, the actual FFT model gives the best performance for all GS asterisms as expected. The Fried model gives worse performance than the Hudgin-like model, even though a slope is defined with more points in the Fried model than the one in the Hudgin-like model. This is because the Fried model is affected by unseen modes such as waffle (see Appendix). In fact, we need to boost a regularization for the Fried model (i.e. assuming larger $\bm{\Sigma_{\eta\eta}}$ in \eref{eq:covIterative1} even without noise) compared to other models to optimize its performance. The sparse method is slightly worse than the Toeplitz with the FFT model due to the sparse approximation of the regularization term $\bm{\Sigma_{\varphi\varphi}}^{-1}\approx\bm{L}^T\bm{L}$ and the limited spatial sampling of the layered phase assumed in the reconstructor, which is set to $d/2$ in this case.

In the analytic WFE, the noise propagation term $\sigma_{noise}$ slightly increases with the asterism radius but its contribution to the total WFE is much smaller than the tomographic WFE ($\sigma^2_{\beta, tomo}=$30\,nm and 50\,nm at asterism radius of 5 and 55\,arcsec, respectively), and there is no clear difference in the noise propagation between the reconstructors. Therefore, the total WFE $\sigma$ in \fref{fig:performance1} is dominated by the tomographic WFE $\sigma_{tomo}$.

The right panel of \fref{fig:performance1} shows the quadratic WFE difference from the Toeplitz method with the FFT gradient model to provide more detail on the performance comparison. The quadratic difference in WFE of the Hudgin and Fried gradient models with respect to the FFT model are almost constant with asterism radius, which are around 35\,nm and 70\,nm, respectively.

\hlone{
In addition to the sparse reconstructor with the approximated regularization $\bm{L}^T\bm{L}$, the sparse one with the non-approximated regularization $\bm{\Sigma_{\phi\phi}}^{-1}$ (of \fref{fig:performance1}) are plotted in \fref{fig:performance1} (the triangles and the dot-dashed line in the right panel) to distinguish the impact of the sparse-approximated regularization and the limited spatial sampling of the layered phase in the sparse reconstructor. The performance of the sparse reconstructor with $\bm{L}^T\bm{L}$ worsens with asterism radius compared to the Toeplitz method with the FFT model, whereas the non-approximated sparse reconstructor are degraded by 25\,nm constantly over the asterism. This indicates that the sparse regularization term that under-regularizes curvature-free modes \cite{lee07} has impact on the tomography performance especially for larger asterisms. On the other hand, the impact of the limited sampling is constant over altitude. If we increase the spatial sampling for the sparse reconstructor to reduce the impact of the limited sampling, the performance of the sparse reconstructor converges to one of the FFT model. However, in this case, the computational advantages of the sparse reconstructor are dimmed because more calculations are required.}

The result for LGSs-based tomography case is shown in \fref{fig:performance2}. Since the non-approximated $\bm{\Sigma_{\phi s}}$ for a LGSs-based case cannot be computed with the FFT gradient model due to the non-uniformity of the spatial sampling of the measurement over altitudes caused by the cone effect, the result from the simulation is plotted for the LGSs case. Although the Hudgin and Fried gradient model are constantly worse than the FFT gradient model as with the NGSs-based case in \fref{fig:performance1}, the WFE with the sparse-based reconstructor with the actual regularization term is better than one of the Toeplitz method with the FFT model. We interpret this as a result of the additional interpolation in the Toeplitz method, $\bm{P_k'}$ in \eref{eq:ToeplitzLGS}, to keep the Toeplitz structure in $\bm{\Sigma_{\phi s}}$ for a LGSs-based case. For asterisms larger than $\sim20$\,arcsec, a larger effect on performance of the approximated regularization term in the sparse reconstructor is observed with the Toeplitz-based reconstructor outperforming the sparse reconstructor under these conditions.

Next, we discuss the robustness of the reconstructor with respect to turbulence conditions. The tomographic error $\sigma^2_{tomo}$ in \eref{eq:analytical3} can be expressed as the linear combination of the tomographic error caused at different altitudes
\begin{align}
    \sigma_{\beta,tomo}^2&=\sum_k C_n^2(h_k)\varepsilon_{\beta,tomo}^2(h_k),
\end{align}
where $\epsilon_{tomo}^2(h_k)$ is a normalized tomographic WFE computed with $C_n^2(h_k)=1$, referred to as \textit{vertical error distribution}
(VED)\cite{gendron14}. The VED corresponds to the estimation capability of the reconstructor as a function of altitude. 

\begin{figure}[b]
    \centering
    \includegraphics[bb=0 0 360 252,width=.44\textwidth]{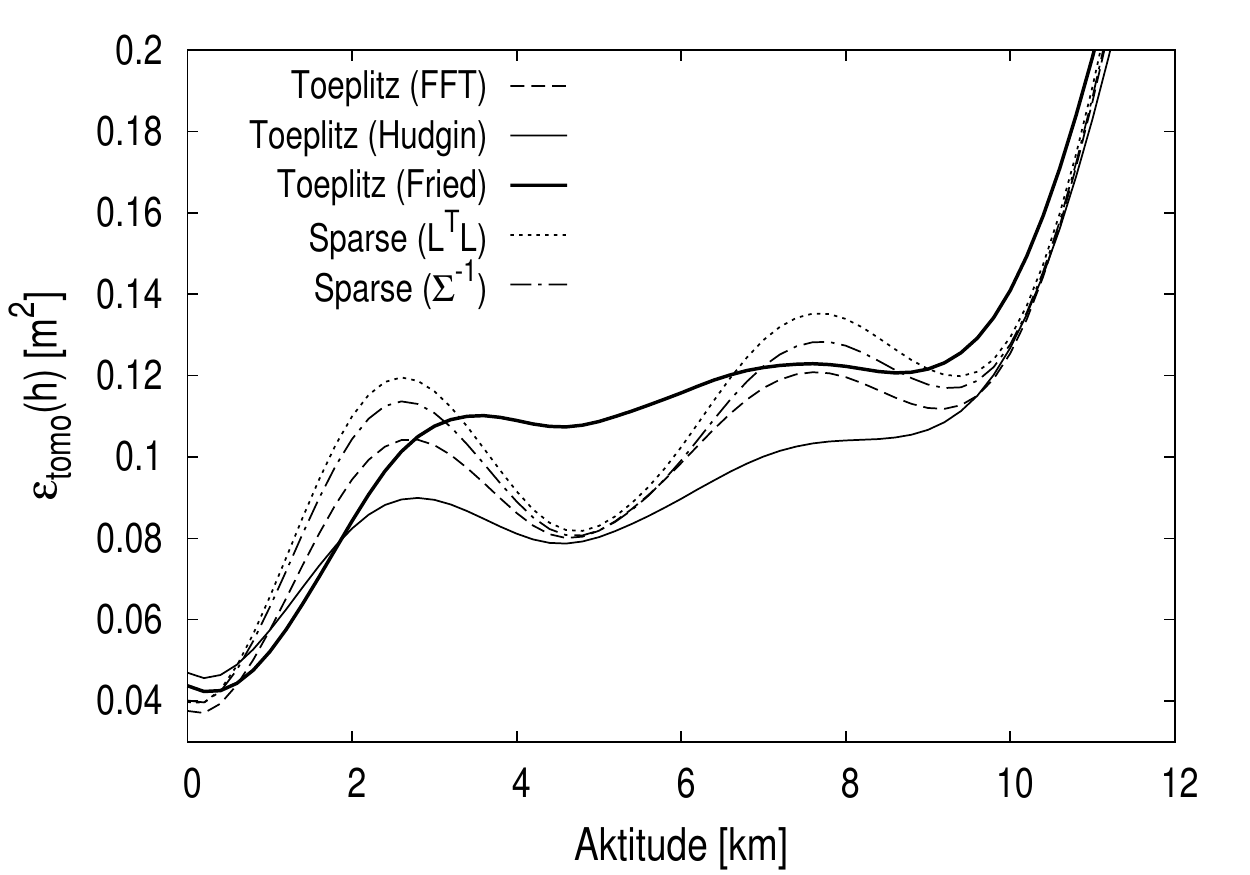}
    \caption{Normalized tomographic WFE as a function of altitude (VED).}
    \label{fig:VED}
\end{figure}

\fref{fig:VED} shows the VED with different gradient models and reconstructors. The reconstructors assume the turbulence model in \tref{tab:simu_param1} and the tomographic WFR is done with three NGSs. There are three hollows at 0, 5 and 10\,km on the VEDs which indicate that the reconstructors, except for the Fried model, reduce aberrations effectively at the altitudes assumed in the reconstructors. At the ground, the Toeplitz method with the FFT model is better than any other reconstructors, and this is the reason why the FFT model gives the best performance in \fref{fig:performance1}. The performances of the Fried model at altitudes larger than 2\,km are poorer than any others because of the unseen waffle mode.

\hlone{
The Toeplitz reconstructor with the Hudgin-like model has the flattest VED profile between the corrected hollows, which means it is the most robust to turbulence layers height variability. Unlike the Fried model, the Hudgin-like model is not affected by the high-spatial frequency waffle mode. In the other words, the Hudgin-like model can intrinsically filter out the problematic modes. We interpret this such as the filtering feature of the Hudgin-like model also works for the unexpected turbulence layers and results in the good robustness of the Hudgin-like model. Conversely, if we boost the regulalization term in the reconstructor, the robustness curve becomes flatter, i.e. the $\varepsilon_{tomo}(h)$ increases at the reconstructed altitudes (0, 5, 10\,km in this case) but decreases in-between.}

The sparse methods are almost as sensitive to the turbulence altitudes as the Toeplitz reconstructor, with further robustness limitations stemming from the coarse spatial sampling of the phase and the regularization.


\section{Performance for HARMONI on the ELT}
\label{sec:eltHarmoni}

We now turn our attention to the case of HARMONI, a visible and
near-infrared (VIS/NIR) integral field spectrograph (IFS), providing
the ELT 37\,m diameter telescope's core spectroscopic capability. It is designed to be
assisted by a LTAO system with 6 LGSs and the deformable M4 mirror on ELT.

\begin{figure}[t]
    \centering
    \includegraphics[bb=0 0 273 324, width=0.4\textwidth]{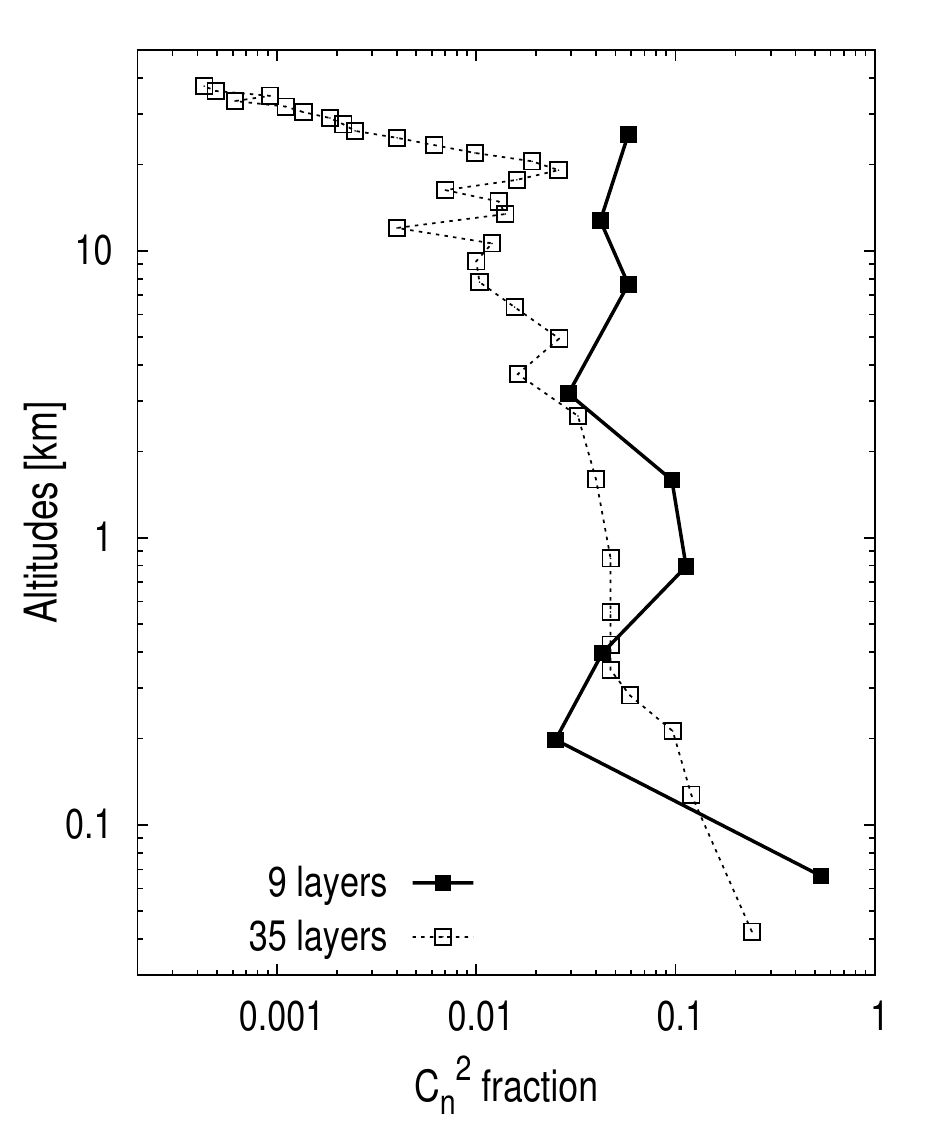}
    \caption{Atmosphere profiles used in the simulation of the HARMONI LTAO system.}
    \label{fig:atmProfile}
\end{figure}

\tref{tab:simu_param2} shows the set up of the HARMONI LTAO system simulations. We assume that the WFSs are noise-free and there is no LGS spot-elongation on the WFS detector. Optimization with respect to the LGS spot elongation for the HARMONI LTAO system will be discussed in the different paper \cite{blanco17}. We assume the DM has actuators with the Fried geometry, which is different from the actual M4 configuration. Two turbulence profiles used for the assessment are shown in \fref{fig:atmProfile}. The Fried parameter is 0.1275\,m at $z=45$\,degrees and the outer scale is 25\,m. The anisoplanatic angles $\theta_0$ for 35 and 9 layers profiles are 2.5\,arcsec and 2.81\,arcsec, respectively, and $\tau_0$ are 8.21\,msec and 8.23\,msec, respectively. The LTAO system is controlled with a split tomographic approach. The tip, tilt and focus are controlled in closed-loop with an on-axis NGS observed by a 2$\times$2 WFSs in H-band. We focus only on the high-order correction here. \hltwo{The low-order modes removal in Section 2.F is taken into account in the reconstruction.}

\subsection{LTAO Performance with different algorithm and iterative solvers}

The convergence property of different iterative solvers is now investigated. The FFT gradient model is used for the Toeplitz method. \fref{fig:convergenceOL} shows the convergence curve of the residual WFE with different cold-started algorithms (i.e. we start from the uncorrected wave-front measurement, not a, hopefully close, previous guess) and open-loop measurements. While the convergence rate of the Toepliz method is independent from the number of reconstructed layers, the sparse method needs a larger number of iterations when the number of estimated layers increases. For the Toeplitz method, the MINRES algorithm is slightly more stable than the BICGSTAB. The CG algorithm doesn't converge with the Toeptliz method, which is not plotted in \fref{fig:convergenceOL}. 
\hltwo{For the sparse algorithm, the CG solver shows the fastest convergence in three solvers followed by BICGSTAB and MINRES solvers. }
In the 35 layers case, the sparse method never achieves the residual WFE given by the Toeplitz method because of the poor convergence of some low-order modes like astigmatisms. 
In the remainder of this paper, MINRES and CG are used for the Toeplitz and sparse methods, respectively. 

\fref{fig:convergencePOL} and \tref{tab:result} show the LTAO performance for a given number of solver iterations evaluated from a long exposure simulation (1000 frames corresponding to 2\,s) using warm start (i.e. iterative solvers start from the previous guess) and pseudo open loop control (POLC). For both of the 9 and 35 layers cases, the WFE curves of the Toeplitz algorithm is almost flat beyond roughly 25 solver iterations, with only a slight performance improvement, less than 1\,nm, from 25 iterations to 50 iterations. On the other hand, the sparse algorithm shows certain performance improvement even with more than 25 iterations. Previous work reported that 30 CG iterations is enough for POLC split tomography with the sparse tomographic algorithm for a 30\,m telescope case \cite{gilles08a}. Although TMT's system complexity is different from our LTAO case, the required number of solver iterations is consistent with each other. The relative performance gains by the Toeplitz method compared with the sparse method are 30\,nm and 54\,nm with 50 solver iterations for 9 and 35 layers case, respectively, in quadratic WFE. 

\begingroup
\renewcommand{\arraystretch}{1.1}
\begin{table}[t]
    \centering
    \begin{tabular}{|l|l|}
    \hline
        \textbf{Telescope} & Aperture Diameter: $D=2R=37$\,m. \\
                    & Central Obstruction: 30\%. \\
                    & Zenith angle: $z=45$\,degree. \\
    \hline
        \textbf{Guide star} &  \\
        High-order & 6 LGSs at $h_{lgs}=$90\,km$/z$. \\
                    & Hexagonal asterism with a radius of $R/h_{lgs}$. \\
                    & assumed as point sources. \\
        Low-order & 1 tip/tilt/focus sensing on-axis NGS. \\
    \hline
        \textbf{WFS}         & \\
        High-order  & 6 Shack-Hartmann WFSs. \\
                    & $74\times74$ subapertures/WFS. \\
                    & subaperture diameter: $d=0.5$\,m. \\
                    & working at 589\,nm. \\
                    & assumed as noiseless.\\
        Low-order   & 1 Shack-Hartmann WFSs. \\
                    & $2\times2$ subapertures/WFS. \\
                    & working at H-band. \\
                    & assumed as noiseless. \\
    \hline
        \textbf{DM} & $75\times75$ Fried geometry DM.\\
                    & Conjugation height: 0\,km\\
    \hline
        \textbf{Atmosphere} & Fried parameter: $r_0=\cos^{5/3}(z)0.157$\,m. \\
                    & Outer scale: $\mathcal{L}_0=25$\,m. \\
                    & Number of layers: $N_{layer}=9, 35$. \\
                    & Altitude and $C_N^2$ fraction shown in. \\
    \hline
        \textbf{Control} & Split control. \\
        High-order  & Pseudo open-loop (POL).\\
                    & POL gain: 0.5\\
                    & Frame rate: 500Hz. \\
        Low-order   & Closed-loop. \\
                    & CL gain: 0.5\\
                    & Frame rate: 500Hz. \\ 
    \hline
    \end{tabular}
    \caption{Parameters for numerical simulation used in Section 4.}
    \label{tab:simu_param2}
\end{table}
\endgroup

\begin{table}[b]
    \centering
    \begin{tabular}{cccc}
        \hline
        & \hspace{1.5mm}\# of solver\hspace{1.5mm} & \hspace{1.5mm}WF\hspace{1.5mm} & \hspace{1.5mm}WFE\hspace{1.5mm}\\
        & iterations & (9 layers) & (35 layers)\\
        &  & [nm] & [nm] \\
        \hline
        Toeplitz & 25 & 164.34 & 174.49 \\
        Toeplitz & 50 & 163.39 & 174.01 \\
        Sparse & 25 & 169.68 & 190.74\\
        Sparse & 50 & 166.17 & 182.32 \\
        \hline
    \end{tabular}
    \caption{Simulation results comparison for the two classes of iterative methods.}
    \label{tab:result}
\end{table}

\subsection{Number of operations for reconstruction}

\begin{figure}
    \centering
    \includegraphics[bb=0 0 360 252,width=0.45\textwidth]{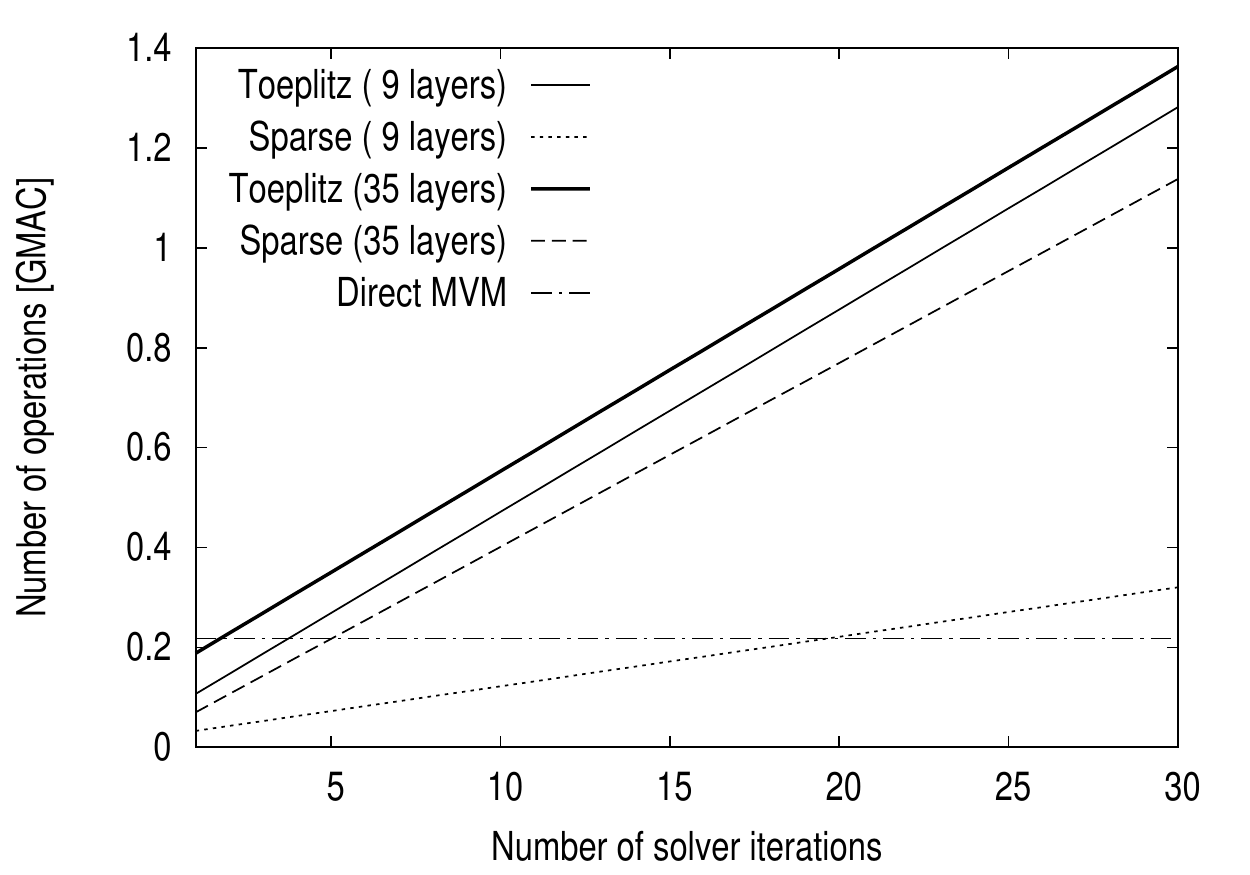}
    \caption{Number of operations in giga MAC as a function of the number of solver iterations. As a comparison, the number of operations for the direct MVM reconstruction is plotted (dash-dotted lines).}
    \label{fig:number}
\end{figure}

The performance shown above in turn is counter-balanced by the actual number of operations
(measured in terms of multiply-and-accumulate (MAC) operations) of the
sparse-based methods. \fref{fig:number} shows such a metric for the
two cases explored above with 9 and 35 estimated layers.

For the computation of the total number of operations, we assume that the number of operations required for one FFT is \hltwo{$5 N\log N$} and that the number of operations for one MVM with a sparse matrix is equal to the number of non-zero elements in the sparse matrix. A MVM for the fitting step is taken into account, where the size of the fitting matrix is (number of valid DM actuators)$\times$(number of reconstructed phase points). 

As $N_{layer}=9$ the computational complexity of the Toeplitz method is larger than one of the sparse method as $N_{layer}=9$. On the other hand, the difference between the Toeplitz method and the sparse method is much smaller than one as $N_{layer}=9$. This indicate that the Toeplitz method is less dependent on the number of reconstructed layers than the sparse method.

\begin{figure*}
    \centering
    \includegraphics[bb=0 0 360 252,width=0.45\textwidth]{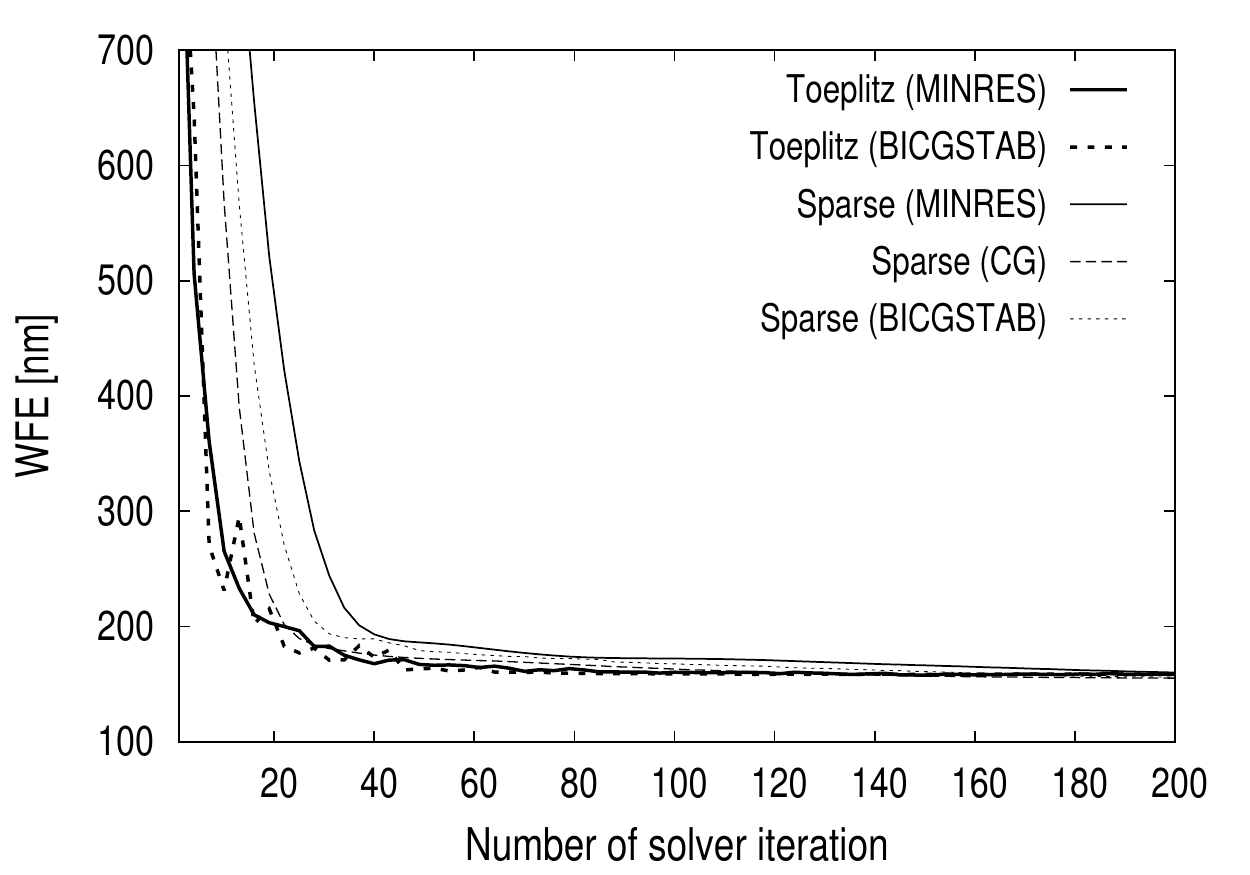}
    \includegraphics[bb=0 0 360 252,width=0.45\textwidth]{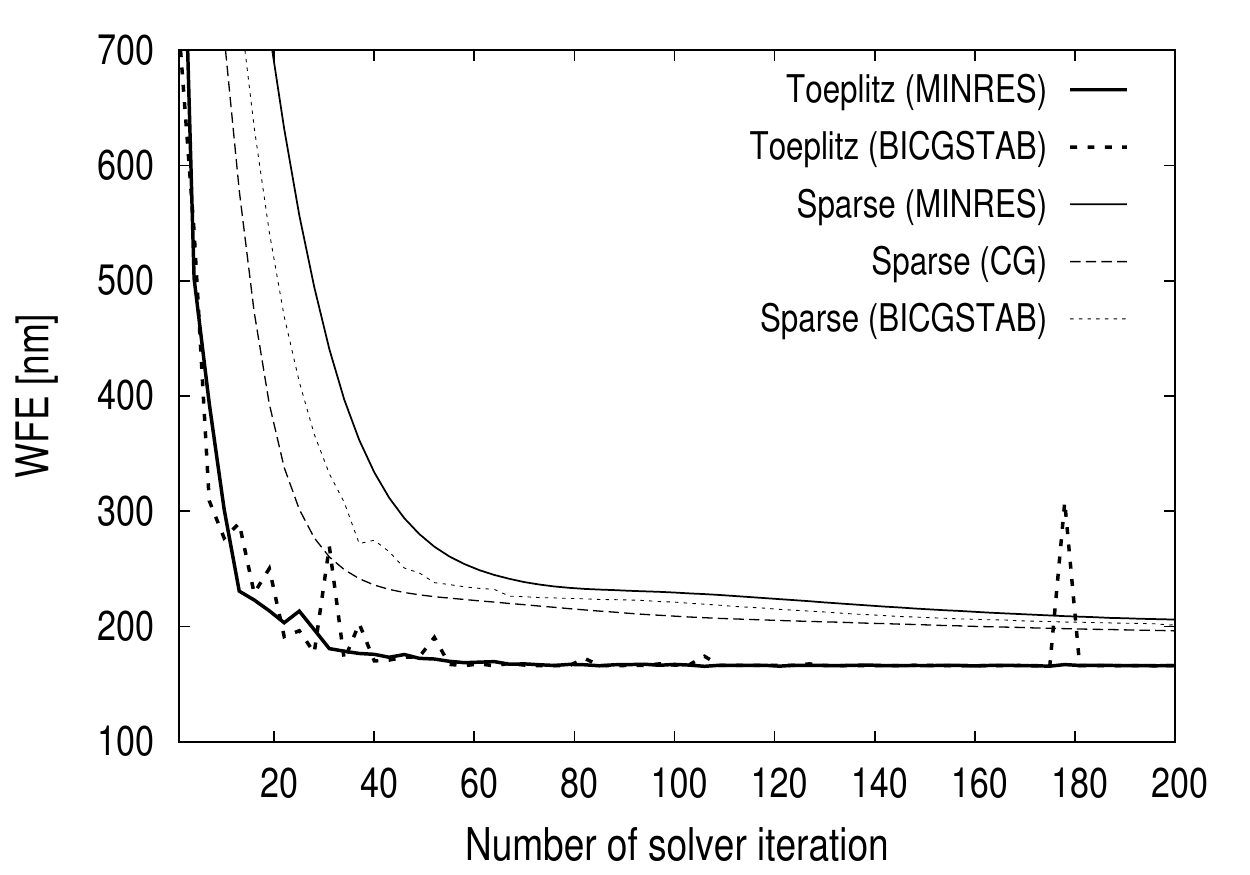}
    \caption{Convergence of different iterative solvers in open-loop reconstructions for 9 (left panel) and 35 layers (right panel) with Toeplitz-based and sparse algorithm: conjugate-gradient (CG), minimum residual method (MINRES) and bi-conjugated gradient stabilized method (BICGSTAB). The FFT gradient model is used for the Toeplitz-based algorithm. The Toeplitz-based method doesn't converge with CG. The iterative solvers are derived with cold start i.e. solver iteration starts from a zero initial vector.}
    \label{fig:convergenceOL}
    \vspace{10mm}
    \centering
    \includegraphics[bb=0 0 360 252,width=0.45\textwidth]{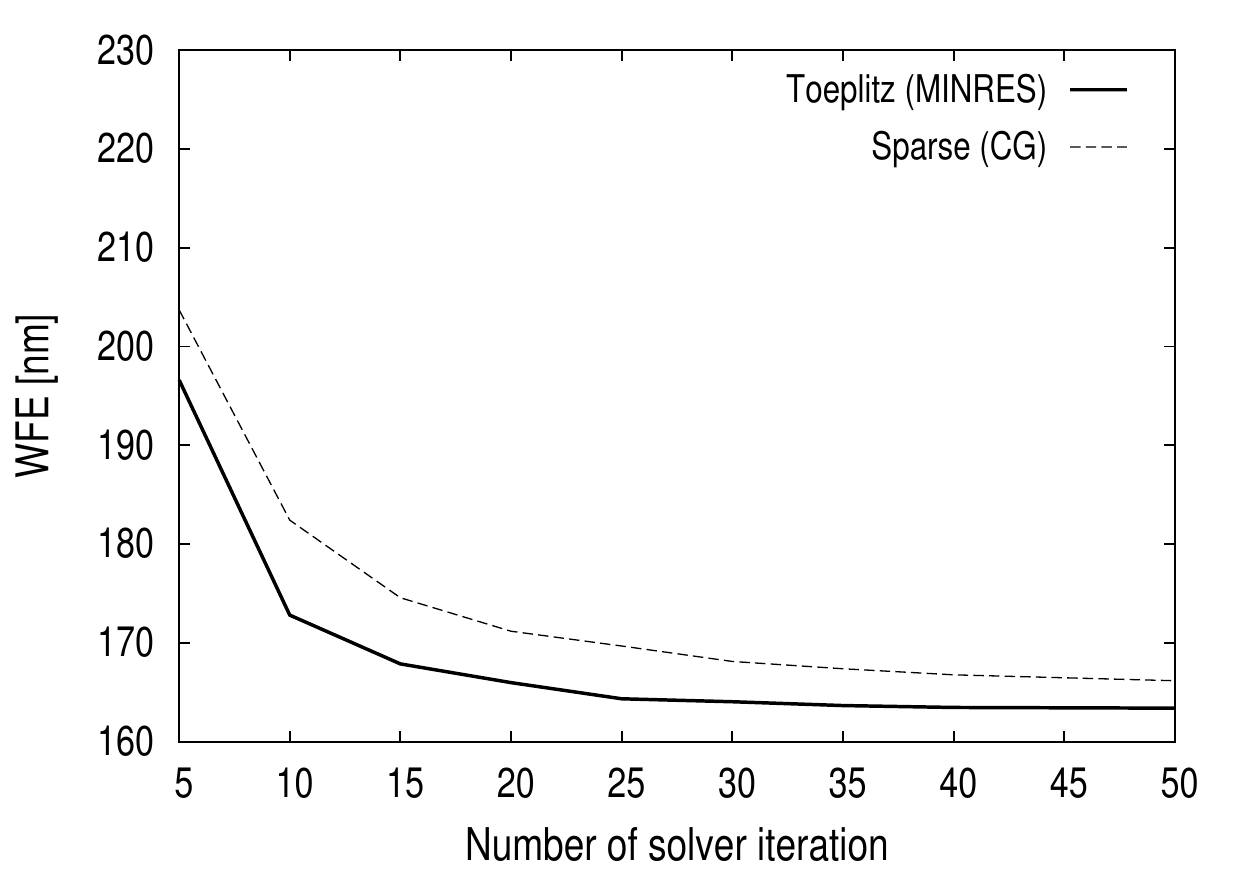}
    \includegraphics[bb=0 0 360 252,width=0.45\textwidth]{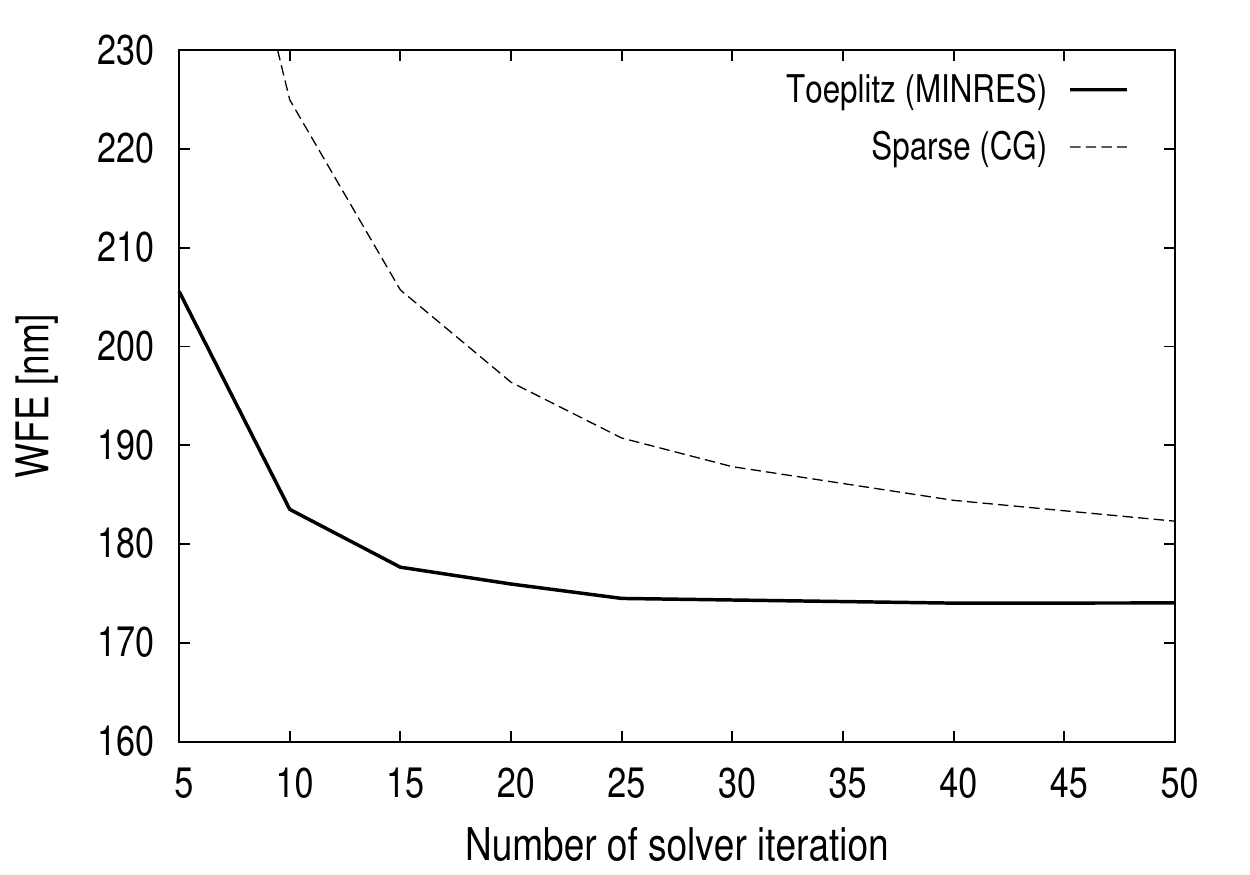}
    \caption{Convergence of iterative solvers in pseudo open-loop reconstructions for 9 (left panel) and 35 layers (right panel). The MINRES and CG solver are used for the Toeplitz and sparse methods, respectively. The gradient model for the Toeplitz method is the FFT model.}
        \label{fig:convergencePOL}
\end{figure*}

As a comparison, we also plot the number of operation for a simplest tomographic reconstruction with a direct MVM, in which case the reconstruction is done by one MVM using a full matrix consisting of the fitting matrix and a pre-computed reconstruction matrix (e.g. computing \eref{eq:covBasedReconstructors1}). It should be noted that the direct MVM reconstruction needs \hltwo{generally} a huge off-line computation to invert a large covariance matrix $\bm{\Sigma_{s_0s_0}}$ with a computational complexity of $(2N_{gs}n)^3$. From a pure computational burden point of view, only 4 solver iterations are allowed for the Toeplitz method to compete with the direct MVM reconstruction in term of the number of operations, which is not enough to maximize LTAO performance according to \fref{fig:convergencePOL}. This however needs to be assessed on actual hardware which exploits locality and memory re-use. 

\clearpage

\section{Real time readiness}
\label{sec:realtime}

MMSE-based iterative reconstructor are rarely used in real-time applications. The main obstacle to these algorithms more widespread use is their iterative nature. A real-time AO system would require the estimation of the wave-front to be done in typically less than 2\,ms. Because an iterative method takes several iterations to converge, meeting the latency requirement of a real-time system with an iterative algorithm is easier said than done. Another difficulty is that the parallelization of iterative algorithms is not as straightforward as a the parallelization of a simple MVM based reconstructor.

For a NGS-based SCAO system, the real-time performance is evaluated by R. Conan \cite{Conan-14} using parallelization with GPU (NVIDIA Tesla M2090) and CUDA. The Toeplitz method can achieve a WFE of 82\,nm with 16 MINRES solver iterations for a SCAO system with a SH-WFS having 84$\times$84 subapertures and $D=42$\,m. It takes 3.75\,ms for one WFR. If the number of MINRES solver iterations is set such the runtime is less than 2\,ms, the achieved WFE becomes slightly worse to 92\,nm with 8 solver iterations.

The Toeplitz method is then applied to a LTAO system on a 25\,m diameter telescope with 6 LGSs evenly located 30\,arcsec off-axis and 6 60$\times$60 SH-WFS. At cold start, where MINRES iteration starts with a zero initial guess, the system converges with 141 iterations in 71 ms to a WFE RMS of 115\,nm. At warm start \cite{lessard08}, the number of iterations is reduced to 70 in 36\,ms. Here the iterative solver uses as initial guess the wave-front 2\,ms before. If the the iterative solver uses as initial guess the wave-front 1\,ms before, then the algorithm converge in 25\,ms with 46 iterations.

In this section, it has been shown that the Toeplitz method is real time ready for some SCAO systems but need further improvements to be usable for LTAO systems.

\section{Conclusion}
\label{sec:conclusion}
We have provided an efficient implementation of an exact tomographic reconstruction method (with respect to a MMSE cost functional) that exploits the Toeplitz nature of the spatio-angular reconstructor formulation. This work expands that of Conan~\cite{Conan-14} to the multi-wave-front, tomographic case using natural and laser guide stars. 

Salient features of spatio-angular reconstructors for NGS LTAO/MOAO/GLAO systems are the independence from the number of estimated layers and the fast convergence rate using the MINRES algorithm. When using LGSs however, extra calculations are needed involving interpolation at pre-defined layer heights to keep the Toeplitz structure of the matrices. This makes the method usable also in MCAO with the caveat that the calculations depend explicitly on number of estimated layers.

Regarding specifically LTAO systems on the European ELT, performance with the Toeplitz algorithm is enhanced by $\sim 60$\,nm rms with improved robustness to altitude variations (Fig. \ref{fig:VED}) with respect to sparse-based tomography. However, the additional interpolation steps needed in the adaptation to the LGS case to conserve the Toeplitz structure involving both planar and spherical wave-propagation (§\ref{sec:mvm_2rtb}) leads to a number of operations after convergence that can be of the order of or greater than that we would realize with a direct MVM reconstruction. Albeit, the covariance matrices can be evaluated very rapidly with a minimal memory footprint. The implementation provided avoids the inversion of any large matrices which is particularly appealing for physical-optics Monte Carlo simulations.  

An optimized implementation on a multi-GPU architecture shows that the Toeplitz method can exploit very efficiently features of this architecture but the number of iterations required for suitable performance is beyond what a real-time system can accommodate to keep up with the time-varying turbulence. At this stage, accelerating the convergence by preconditioning the system of linear equations remains a challenge: we hope interested readers can develop the necessary means to further accelerate the algorithm's convergence rates.

\hltwo{Although not discussed in this paper another possible application of the Toeplitz method is to compute the reconstructor $\bm{R_{\varphi_\beta}}$ offline by solving $\bm{\Sigma_{s_\alpha s_\alpha}}\bm{R_{\varphi_\beta}}=\bm{\Sigma_{\varphi_\beta s_\alpha}}$ one column at a time using the 2RBT MVM. It needs to be investigated whether this method meets the offline update rate requirements (e.g. 10\,s or so).
This matrix inversion with the Toeplitz method can also be applied to the analytic evaluation (solving \eref{eq:analytical2}) for large-scale systems.}

\section*{Acknowledgments}
The research leading to these results received the support of Grant-in-Aid for JSPS Fellows (15J02510) and the A*MIDEX project (no. ANR-11- IDEX-0001- 02) funded by the ”Investissements dAvenir” French Government program, managed by the French National Research Agency (ANR). 

\newpage
\appendix
\section{Theoretical Derivation of Covariance Matrix}
\label{sec:Appendix1}

\subsection{FFT slope model}

In this Appendix, we explain theoretical models for SH-WFS gradient measurement. SH-WFSs with $n\times n$ subapertures provide phase gradients averaged over a subaperture centred at $\bm{r_i}$ i.e.
\begin{align}
    s_x(\bm{r_i})&=\frac{\lambda}{2\pi d^2}\int\frac{\partial\phi}{\partial x}(\bm{u})\Pi\left(\frac{\bm{u}-\bm{r_i}}{d}\right)\bm{du},
\end{align}
where $\Pi(x)$ is a rectangular function which is 1 for $|x|\leq1/2$ and 0 otherwise, $d$ is a subaperture diameter and $\lambda$ is an observing wavelength.

A covariance of two slopes with a separation $\bm{r}=\bm{r_j}-\bm{r_i}$ on the pupil is given by
\begin{align}
    \Sigma_{s_xs_x}(\bm{r})&=\langle s_x(\bm{r_i})s_x(\bm{r_j})\rangle\notag\\
    &\begin{aligned}
        &=\left(\frac{\lambda}{2\pi d^2}\right)^2
        \int\hspace{-4pt}\int
        \left\langle
        \frac{\partial\phi}{\partial x}(\bm{u})
        \frac{\partial\phi}{\partial x}(\bm{v})
        \right\rangle\\
        &\hspace{15pt}\Pi\left(\frac{\bm{u}-\bm{r_i}}{d}\right)\Pi\left(\frac{\bm{v}-\bm{r_i}}{d}\right)\bm{du}\bm{dv}.
    \end{aligned}
\end{align}
By considering the inverse Fourier transform of $\Sigma_{ss}(\bm{r_j}-\bm{r_i})$, we get the slope-slope covariance for a single layer at altitude $h_k$ with
\begin{equation}\label{eq:ss_fft}
    \begin{aligned}
        \Sigma_{s_xs_x,k}(\bm{r})=&\mathcal{F}\left\{\lambda^2f_xf_x\Phi_\phi(\bm{f})H^2(\bm{f})\right.\\
        &\left.\times\exp[-2i\pi h_k(f_x\delta\alpha_x+f_y\delta\alpha_y)]\right\}(\bm{r}),
    \end{aligned}
\end{equation}
where $\bm{f}=(f_x,f_y)$ is a spatial frequency, $\Phi_\phi(\bm{f})$ is a phase power spectral density and
\begin{equation}
    H(\bm{f})=\text{sinc}(df_x)\text{sinc}(df_y).
\end{equation}
The phase power spectral density $\Phi_\phi(\bm{f})$ of the von-Karman power spectrum is given with the Fried parameter $r_0$ and the outer scale $\mathcal{L}_0$ as
\begin{equation}
    \Phi_\phi(\bm{f})=0.029r_0^{-5/3}\left(||\bm{f}||^2+\frac{1}{\mathcal{L}_0^2}\right)^{-11/6}.
\end{equation}
The last exponential term in \eref{eq:ss_fft} represents the shift of the pupil projected at $h_k$ due to the angular separation $\bm{\delta\alpha}=(\delta\alpha_x,\delta\alpha_y)$ of two guide stars in the Fourier domain. A covariance value induced by the multiple layers is a sum of the single layer covariances. The rest of the slope-slope covariances, $\Sigma_{s_ys_y,k}$ and $\Sigma_{s_xs_y,k}=\Sigma_{s_ys_x,k}$, are obtained by replacing $f_xf_x$ into $f_yf_y$ and $f_xf_y$ in \eref{eq:ss_fft}, respectively.

With the same way, the phase-slope covariance for a single layer is given by
\begin{align}\label{eq:ps_fft}
    \Sigma_{\phi s_x,k}(\bm{r})&=\langle \phi(\bm{r_i})s_x(\bm{r_j})\rangle\notag\\
    &\begin{aligned}
        &=\mathcal{F}\left\{-i\lambda f_x\Phi_\phi(\bm{f})H(\bm{f})\right.\\
        &\left.\times\exp[-2i\pi h_k(f_x\delta\alpha_x+f_y\delta\alpha_y)]\right\}(\bm{r}),
    \end{aligned}
\end{align}
and $\Sigma_{\phi s_y,k}$ is obtained by replacing the first $f_x$ into $f_y$ in \eref{eq:ps_fft}. 

\hlone{
Since there is no discrete assumption, the FFT model provides an accurate covariance value. However, in actual computation, the FFT covariance model is computed by the FFT, and its actual accuracy depends on the sampling of the FFT. In order to get better performance, this model needs more sampling in the FFT, i.e. requires more computations, which will be relatively heavy computation especially for future ELTs cases.}

\begin{figure}[t]
    \centering
    \includegraphics[bb=0 0 275 216,width=0.32\textwidth]{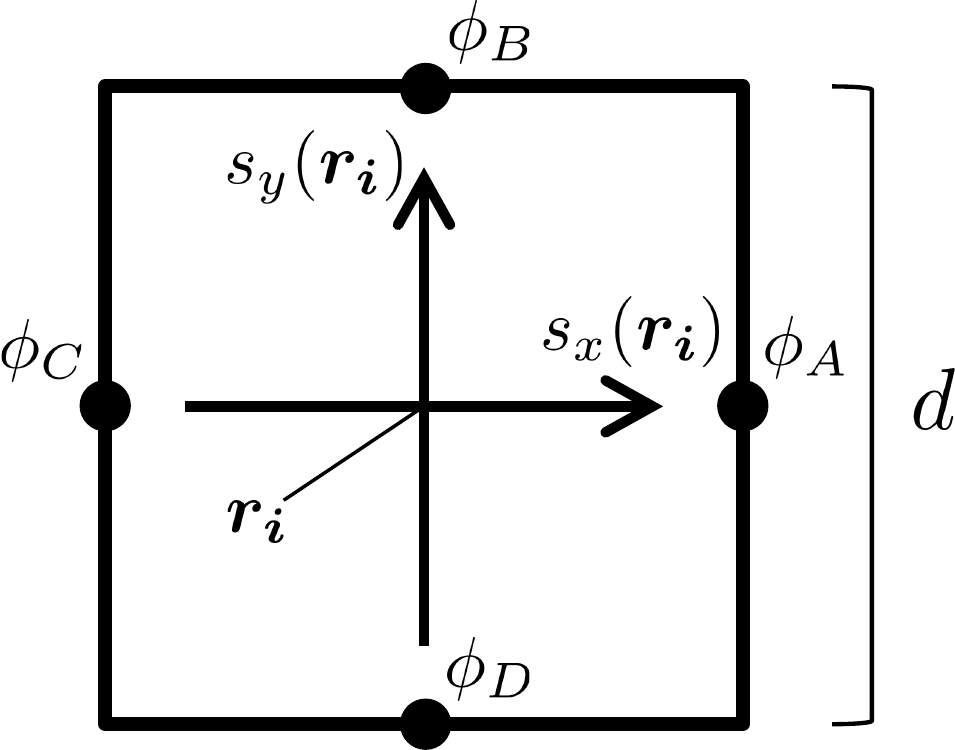}
    \caption{Hudgin-like slope model}\label{fig:slopeModel:hudgin}
\end{figure}

\subsection{Hudgin-like model}
Although the FFT slope model provides an accurate covariance value, its computation is relatively heavy especially for future ELTs cases. In order to reduce the computational burden to compute the covariance matrices, two discrete-approximated slope model have been proposed.

One of the approximated slope models is developed in \cite{Martin-12}, referred to as Hudgin-like slope model in this paper. As shown in \fref{fig:slopeModel:hudgin}, a slope approximated as a phase difference between two points on the edges of the subaperture as
\begin{equation}
    \begin{aligned}
        &s_x(\bm{r_i})=\frac{\lambda}{2\pi d}(\phi_A-\phi_C)\\
        &s_y(\bm{r_i})=\frac{\lambda}{2\pi d}(\phi_B-\phi_D),
    \end{aligned}
\end{equation}
and 
\begin{align}
    \begin{aligned}
        \phi_A=\phi\left(\bm{r_i}+\frac{\bm{d_x}}{2}\right),
        \ \phi_B=\phi\left(\bm{r_i}+\frac{\bm{d_y}}{2}\right),\\
        \phi_C=\phi\left(\bm{r_i}-\frac{\bm{d_x}}{2}\right),
        \ \phi_D=\phi\left(\bm{r_i}-\frac{\bm{d_y}}{2}\right),
    \end{aligned}
\end{align}
where $\bm{d_x}=(d,0)$ and $\bm{d_y}=(0,d)$.

Considering the definition of the phase structure function
\begin{equation}
    D_\phi(\bm{r})=\langle[\phi(\bm{x})-\phi(\bm{x}+\bm{r})]^2\rangle
\end{equation}
and the equality of $2(A-a)(B-b)=-(A-B)^2+(A-b)^2+(a-B)^2-(a-b)^2$, we have a slope-slope covariance of the Hudgin-like slope model for a single layer at $h_k$ as
\begin{equation}\label{eq:hudgin_ss}
    \begin{aligned}
        &\Sigma_{s_x s_x,k}(\bm{r})=\frac{1}{2}
        \left(\frac{\lambda}{2\pi d}\right)^2\times\\
        &[
        -2D_\phi(\bm{\Delta_k})
        +D_\phi(\bm{\Delta_k}+\bm{d_x})
        +D_\phi(\bm{\Delta_k}-\bm{d_x})
        ],
    \end{aligned}
\end{equation}
where $\bm{\Delta_k}=\bm{r}+h_k\bm{\delta\alpha}$,
\begin{equation}
    \begin{aligned}
        &\Sigma_{s_x s_y,k}(\bm{r})=\Sigma_{s_y s_x,k}(\bm{r})=\frac{1}{2}
        \left(\frac{\lambda}{2\pi d}\right)^2\\
        &\times\left[
        D_\phi(\bm{\Delta_k}+\frac{\bm{d_x}}{2}+\frac{\bm{d_y}}{2})
        -D_\phi(\bm{\Delta_k}-\frac{\bm{d_x}}{2}+\frac{\bm{d_y}}{2})
        \right.\\
        &\left.
        -D_\phi(\bm{\Delta_k}+\frac{\bm{d_x}}{2}-\frac{\bm{d_y}}{2})
        -D_\phi(\bm{\Delta_k}-\frac{\bm{d_x}}{2}-\frac{\bm{d_y}}{2})
        \right]
    \end{aligned}
\end{equation}
and $\Sigma_{s_x s_x,k}(\bm{r})$ is got by replacing $\bm{d_x}$ into $\bm{d_y}$ in \eref{eq:hudgin_ss}.

The phase-slope covariance is given by
\begin{equation}\label{eq:hudgin_ps}
    \begin{aligned}
        \Sigma_{\phi s_x,k}(\bm{r})&=
        \frac{\lambda}{4\pi d}\\
        \times&\left[
        D_\phi\left(\bm{\Delta_k}-\frac{\bm{d_x}}{2}\right)
        -D_\phi\left(\bm{\Delta_k}+\frac{\bm{d_x}}{2}\right)
        \right]
    \end{aligned}
\end{equation}
and $\Sigma_{\phi s_x,k}(\bm{r})$ is got by replacing $\bm{d_x}$ into $\bm{d_y}$ in \eref{eq:hudgin_ps}.

The structure function is given for the von-Karman turbulence \cite{Butterley-06} by
\begin{equation}
    \begin{aligned}
        D_\phi(\bm{r})=&0.17253\left(\frac{\mathcal{L}_0}{r_0}\right)^{5/3}\\
        &\times\left[1-\frac{2\pi^{5/6}}{\Gamma(5/6)}\left(\frac{r_0}{\mathcal{L}_0}\right)^{5/6}
        K_{5/6}\left(\frac{2\pi r}{\mathcal{L}_0}\right)\right],
    \end{aligned}
\end{equation}
where $K$ is a modified Bessel function of the second kind.
\begin{figure}[b]
    \centering
    \includegraphics[bb=0 0 235 216,width=0.27\textwidth]{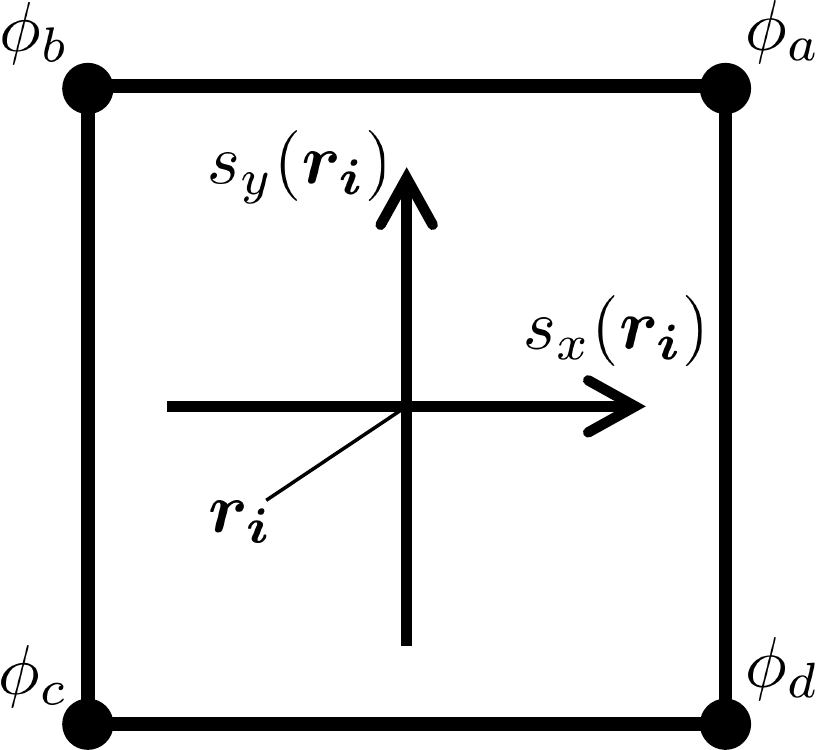}
    \caption{Fried slope model}\label{fig:slopeModel:fried}
\end{figure}

\hlone{
The Hudgin-like covariance model can be computed efficiently compared to the FFT covariance model, but a slope is model by only two discrete points and it is reported that the auto-covariance value of the Hudgin-like model is slightly smaller than a value of the FFT model \cite{Martin-12}. This difference would affect the WFR performance.}

\subsection{Fried model}
Another approximated slope model is defining a slope with 4 phase points on the corners of the subaperture as shown in \fref{fig:slopeModel:fried}, and this is will-known Fried slope model,
\begin{equation}
    \begin{aligned}
       s_x(\bm{r_i})=\frac{\lambda}{2\pi d}\frac{(\phi_a-\phi_b-\phi_c+\phi_d)}{2}\\
       s_y(\bm{r_i})=\frac{\lambda}{2\pi d}\frac{(\phi_a+\phi_b-\phi_c-\phi_d)}{2}
    \end{aligned}
\end{equation}
and
\begin{align}
    \begin{aligned}
        &\phi_a=\phi\left(\bm{r_i}+\frac{\bm{d_x}}{2}+\frac{\bm{d_y}}{2}\right),
        \phi_b=\phi\left(\bm{r_i}-\frac{\bm{d_x}}{2}+\frac{\bm{d_y}}{2}\right),\\
        &\phi_c=\phi\left(\bm{r_i}-\frac{\bm{d_x}}{2}-\frac{\bm{d_y}}{2}\right),
        \phi_d=\phi\left(\bm{r_i}+\frac{\bm{d_x}}{2}-\frac{\bm{d_y}}{2}\right).
    \end{aligned}
\end{align}\par
With the similar computation to the Hudgin-like slope model, we have slope-slope covariance values of the Fried slope model as
\begin{align}
    &\begin{aligned}
        \Sigma_{s_x s_x,k}(\bm{r})&=
        \frac{1}{2}\left(\frac{\lambda}{4\pi d}\right)^2\\
        \times&\left[-4D_\phi(\bm{\Delta_k})+2F_1-2F_2+F_3+F_4\right]
    \end{aligned}\\
    &\begin{aligned}
        \Sigma_{s_y s_y,k}(\bm{r})&=
        \frac{1}{2}\left(\frac{\lambda}{4\pi d}\right)^2\\
        \times&\left[-4D_\phi(\bm{\Delta_k})-2F_1+2F_2+F_3+F_4\right]
    \end{aligned}\\
    &\Sigma_{s_x s_y,k}(\bm{r})=
    \frac{1}{2}\left(\frac{\lambda}{4\pi d}\right)^2
    (F_3-F_4),
\end{align}
where
\begin{equation}
    \begin{aligned}
        F_1&=D_\phi(\bm{\Delta_k}+\bm{d_x})+D_\phi(\bm{\Delta_k}-\bm{d_x})\\
        F_2&=D_\phi(\bm{\Delta_k}+\bm{d_y})+D_\phi(\bm{\Delta_k}-\bm{d_y})\\
        F_3&=D_\phi(\bm{\Delta_k}+\bm{d_x}+\bm{d_y})+D_\phi(\bm{\Delta_k}-\bm{d_x}-\bm{d_y})\\
        F_4&=D_\phi(\bm{\Delta_k}+\bm{d_x}-\bm{d_y})+D_\phi(\bm{\Delta_k}-\bm{d_x}+\bm{d_y}).
    \end{aligned}
\end{equation}
The phase-slope covariances are given by
\begin{align}
    \Sigma_{\phi s_x,k}(\bm{r})&=
    \frac{\lambda}{8\pi d}(-F_5+F_6+F_7-F_8),\\
    \Sigma_{\phi s_y,k}(\bm{r})&=
    \frac{\lambda}{8\pi d}(-F_5-F_6+F_7+F_8),
\end{align}
and
\begin{equation}
    \begin{aligned}
        &F_5=D_\phi\left(\bm{\Delta_k}+\frac{\bm{d_x}}{2}+\frac{\bm{d_y}}{2}\right)\\
        &F_6=D_\phi\left(\bm{\Delta_k}-\frac{\bm{d_x}}{2}+\frac{\bm{d_y}}{2}\right)\\
        &F_7=D_\phi\left(\bm{\Delta_k}-\frac{\bm{d_x}}{2}-\frac{\bm{d_y}}{2}\right)\\
        &F_8=D_\phi\left(\bm{\Delta_k}+\frac{\bm{d_x}}{2}-\frac{\bm{d_y}}{2}\right)
    \end{aligned}
\end{equation}

\hlone{
This model also provides the faster covariance computation than the FFT model. Compared to the Hudgin-like model, the Fried model needs more computation than the Hudgin-like model, but would give more realistic value because a slope is defined by four points instead of two points. However, one concern about the Fried model is that the Fried slope model suffers from an unseen mode with high spatial frequency, so-called \textit{waffle error} or \textit{checker board error} \cite{hardy98,correia14a}. This error can be constrained by regularization, but has an impact on WFR.}


\end{document}